\documentclass[preprint,nofootinbib,amssymb]{revtex4}
\usepackage{bbding,multirow,subfigure}
\usepackage{graphicx}
\newcommand{\comment}[1]{}
\pdfoutput=1
\begin{document}
 
\title{Aether Unleashed}

\author{Cristian Armendariz-Picon}
\affiliation{Department of Physics, Syracuse University, NY 13244-1130, USA.\\
Departament de F\'isica Fonamental i Institut de Ci\`encies del Cosmos, Universitat de Barcelona, 080193 Barcelona, Spain.}

\author{Alberto Diez-Tejedor}
\affiliation{Instituto de Ciencias Nucleares, Universidad Nacional Aut\'onoma de M\'exico, M\'exico D.F. 04510, M\'exico. \\
Dpto. de F\'isica Te\'orica, Universidad del Pa\'is Vasco,  48080 Bilbao, Spain.
}

\begin{abstract}
We follow a low-energy effective theory approach to identify  the general class of  theories that describes a real vector field (of unconstrained norm) coupled to gravity. The resulting set may  be regarded as a generalization of the conventional vector-tensor theories, and as a high-momentum completion of aether models.  We study the conditions that a viable cosmology, Newtonian limit and absence of classical and quantum instabilities impose on the parameters of our class of models, and compare these constraints with those derived in previously studied and  related cases.  The most stringent conditions arise from the quantum stability of the theory, which allows dynamical cosmological solutions only for a  non-Maxwellian kinetic term. The gravitational constant in the Newtonian limit turns to be scale dependent, suggesting connections to dark matter and degravitation. This class of theories has a very rich gravitational phenomenology, and offers an ample but simple testing ground to study modifications of gravity and their cosmological implications.

\end{abstract}
\maketitle

\section{Introduction}
The symmetries of a spacetime severely constrain the transformation properties of its matter content. In Minkowski spacetime for instance Lorentz symmetry allows scalar fields to have a non-vanishing expectation value, forbidding non-scalar fields from being non-zero. It is perhaps because of these constraints that most cosmologists have focused on  cosmological solutions with scalar fields.

Nevertheless, in an expanding universe no symmetry principle prevents the existence of non-vanishing tensor fields, as long as  they are constant in space and invariant under \emph{spatial} rotations.   Since only bosonic fields can acquire large occupation numbers, and if we restrict ourselves to fields of spin less than two, this leaves vector fields (or one-forms) as essentially the only additional type of  field that may be relevant cosmologically.  If the vector field only couples to gravity, the presence of a preferred Lorentz frame will remain  unobservable in the matter sector, while, of course, Lorentz symmetry is broken anyway by the expansion of the universe and is not of particular relevance on large scales. 

Massless vector-tensor theories were considered long time ago  to explore different modifications of gravity, and as alternatives to the more prevalent scalar-tensor theories at that time \cite{Will:1972zz, Hellings:1973zz}. Indeed, as we shall see, vector-tensor theories have a much richer  gravitational phenomenology than  scalar-tensor theories, and thus offer a wider framework to study departures from general relativity.  Theories with vectors have also experienced a modest revival in connection with inflation and the accelerated expansion of the universe \cite{Ford:1989me, ArmendarizPicon:2004pm,Wei:2006tn,Kanno:2006ty,Novello:2006ng,Boehmer:2007qa,Koivisto:2007bp,Jimenez:2008au,Koivisto:2008ig,Golovnev:2008cf,Koivisto:2008xf,Yokoyama:2008xw,Kanno:2008gn,Watanabe:2009ct,Koh:2009ne} and studies  of the spontaneous breaking of Lorentz invariance in the so-called Einstein-aether models of Jacobson and Mattingly \cite{Jacobson:2000xp} (see \cite{Gasperini:1987nq} for earlier discussions, and \cite{Jacobson:2008aj} for a status report.) 
 
In this article we study a class of theories that somewhat differs  from previous models. We explore the most general vector-tensor theory expected to describe our universe at low energies, large distances and long timescales.  We thus follow a low energy effective action approach and consider all generally covariant terms with the lowest number of vector field derivatives. Since we deal with massive vectors, a consistent quantum description at low energies  does not require gauge invariance,  so  many different kinetic terms for the vector field  are possible. Conversely, if we do not assume gauge invariance, no symmetry principle prevents the appearance of  mass terms for the vector, or vector self-interactions for that matter. Our effective action thus contains all generally covariant terms with two vector field derivatives and a general quartic potential\footnote{As we shall see, the ``potential" $V$ actually contains kinetic terms for some of the vector degrees of freedom.} $V$ that general covariance forces to depend on the squared norm of the vector, $A_\mu A^\mu$.  

Up to a constant term, any scalar quartic potential  can be  written in the form
\begin{equation}\label{eq:potential}
	V(A_\mu A^\mu)=\lambda (A_\mu A^\mu+ M^2)^2,
\end{equation}
where $M$ is a mass scale and $\lambda$ a dimensionless parameter.  In fact, around an extremum, any potential can be locally approximated by (\ref{eq:potential}). Such a dependence opens up a quite interesting possibility, because the vacuum of the theory has a non-vanishing vector field $A_\mu A^\mu=-M^2$, and thus in flat space Lorentz symmetry is spontaneously broken.\footnote{We consider here positive values of $M^2$. In that case, $A^\mu$ is timelike in the broken phase. If $M^2$ is negative, the vector vev  is spacelike.}  In fact, in some modern approaches to quantum
gravity Lorentz invariance is expected to be broken at high energies \cite{Kostelecky:1988zi,Horava:2009uw}. However, if broken, it should be done in a very subtle way. If the breaking relies in the existence of a fundamental  preferred frame at the Planck-scale, some unnatural fine tuning issues appear \cite{Collins:2004bp}, or additional ill-behaved modes appear in the spectrum \cite{Charmousis:2009tc,Blas:2009yd}. Moreover, a fixed external structure goes drastically against our currently
accepted principles (i.e. general covariance). This suggests that, probably, the more reasonable way in which Lorentz invariance could be broken is by the existence of a vacuum expectation value (vev) for some non-scalar field, which, as in the ghost condensate \cite{ArkaniHamed:2003uy}, may be the gradient of a scalar.

A widely studied class of models in which Lorentz invariance is spontaneously broken involves  the Einstein-aether models introduced by Jacobson and Mattingly in \cite{Jacobson:2000xp}. One of the most distinctive properties of these models is that the norm of the vector field is constrained to have a constant value.  This constraint may appear artificial, but it could be naively justified as the limit in which the coupling constant $\lambda$ in the potential of equation (\ref{eq:potential}) tends to infinity. Hence, aether models may be regarded as a particular case of the class of theories we consider here.  Because we drop  the fixed-norm constraint we call this class of theories ``unleashed aether" models. Similar kind of actions
have been considered for instance in \cite{Bluhm:2004ep,Bailey:2006fd,Bluhm:2007bd,Kostelecky:2009zr} under the name of ``bumblebee models." In general, the latter are models in which a vector field acquires a vev due to the presence of a potential. In that sense, the unleashed aether is a bumblebee model.  The real difference between the unleashed aether and the most commonly studied bumblebee models lies in the parameters in our action that are assumed to be zero in the bumblebee case. 

The main goal of this article is to study the conditions that self-consistency and phenomenological viability imposes on the parameters  of our class of theories, and to determine to what extent these differ from massless vector fields and aether models. In Section \ref{sec:generalities} we present and discuss general properties of our class of models. In Section \ref{sec:cosmology} we study the properties of cosmological solutions at early and late times, and derive basic constraints on the value of the scale $M$ in the potential. In Section \ref{sec:PPN} we study the Newtonian limit of these theories, and explore their relation with massless theories and aether models. Finally, in Section \ref{sec:stability}, we derive the conditions that classical and quantum-mechanical stability places on the parameter space of unleashed aether models. 

\section{Generalities}\label{sec:generalities}

\begin{figure}
\begin{picture}(220,200)
	\includegraphics[height=8cm]{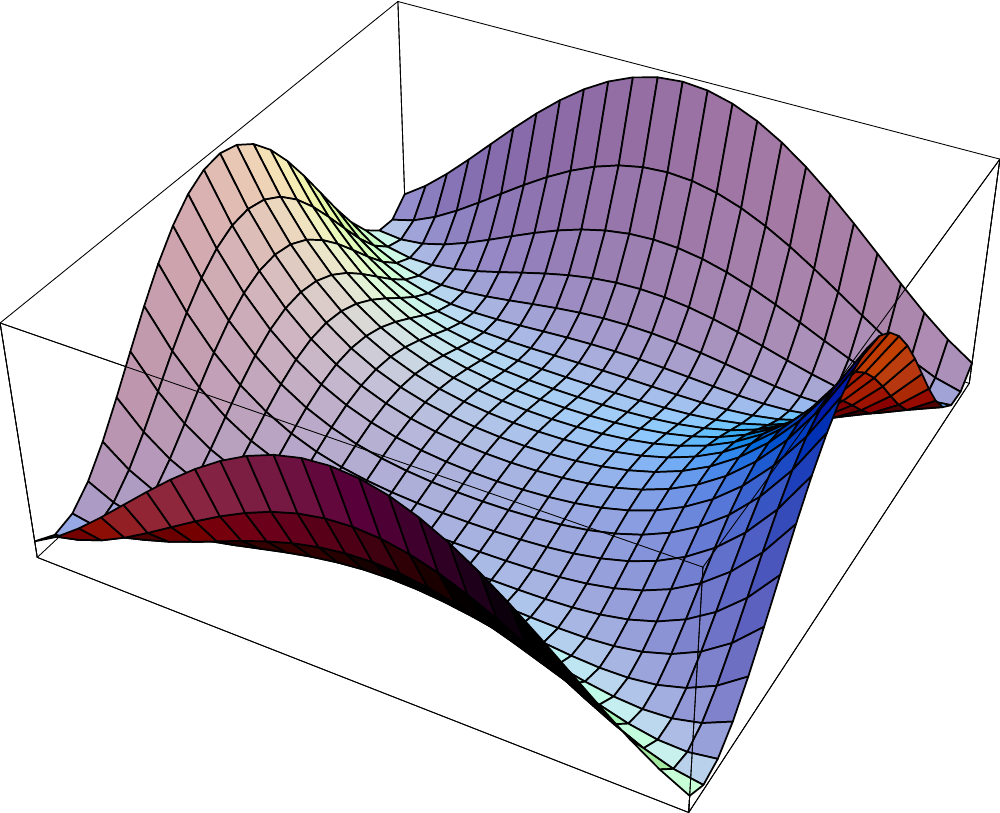}
\put(-200,18){$A_{i}$}
\put(-33,50){$A_{0}$}
\put(-296,100){$V$}
\end{picture}
\label{} 
\caption{A plot of the potential (\ref{eq:potential}) for $M^2>0$ and $\lambda>0$. Because of the indefinite signature of spacetime, the potential does not have the shape of a Mexican hat. Instead, around the two  local minima along the $A_0$ direction, the potential looks like a horse saddle and thus has negative Gaussian curvature. If $M^2<0$ the roles of $A_0$ and $A_i$ are reversed.}
\label{fig:potential}
\end{figure}

We want to identify the most  general  action that describes the dynamics of a real vector field $A^\mu$ coupled to gravity.  Without additional restrictions this action will contain all possible terms compatible with the symmetries of the theory, which we assume to consist just of general covariance and a $\mathbb{Z}_2$ symmetry $A^\mu\to -A^\mu$. If we are interested in low energies however,    we expect only those terms with the least number of derivatives to be relevant in the limit of long distances and times.  We can thus concentrate on the most general generally covariant action with at most two derivatives acting on  the metric and the vector. In principle, such an action would also contain  arbitrary   powers of the vector field, but since a vector has dimensions of mass, terms with too many powers of $A^\mu$ must be suppressed by corresponding powers of an energy scale $\Lambda$. If the background value of the field is smaller than $\Lambda$, the dominant terms are hence given by those with no suppression, that is, by the operators of dimension four. Therefore, we just have to consider the most general action with two derivatives acting on the metric and  vector field, but with no vector field operators of dimension higher than four:  
 \begin{eqnarray}\label{eq:action}
	S=\int d^{4}x \, \sqrt{-g}\Bigg[\frac{R}{16\pi G}
	-\frac{\beta_1}{2} F_{\mu\nu} F^{\mu\nu}-\beta(\nabla_{\mu}A^{\mu})^{2} 
	&+&\beta_{13}R_{\mu\nu}A^{\mu}A^{\nu}+\nonumber \\
	&+&\beta_4 R A_\mu A^\mu-V(A_\mu A^\mu)\Bigg],
\end{eqnarray}
where the $\beta_i$ are dimensionless coefficients, $F_{\mu\nu}\equiv \partial_\mu A_\nu-\partial_\nu A_\mu$,  and $V$ is a  quartic potential of the form (\ref{eq:potential}), which we plot in Figure \ref{fig:potential}.  This is the most general action with two derivatives acting on $A^\mu$ because a  term proportional to $\nabla_\mu A_\nu \nabla^\nu A^\mu$ can be eliminated integrating by parts and using the identity ${[\nabla_\mu,\nabla_\nu]A^\rho=R_{\mu\nu}{}^\rho{}_\sigma A^\sigma}$. The only dimension four operator with one vector derivative, $A_\mu A^\mu \nabla_\nu A^\nu$, is excluded by the discrete symmetry $A^\mu\to -A^\mu$. Observe that we do not assume gauge invariance, which would only allow the Maxwell term in the vector sector. Although gauge invariance seems to be necessary for a consistent description of \emph{massless} vector field interactions \cite{Weinberg:gauge}, it is not required for massive vector fields, at least at sufficiently low energies. The  stability of the theory and phenomenological considerations then set constraints on the values of the so far arbitrary coefficients $\beta_i$ and the derivatives of the potential $V$. When we explore these constraints in Section \ref{sec:stability}, we shall  see that the term proportional to the Maxwell Lagrangian describes a vector field (under Lorentz-transformations), whereas the term proportional to the divergence squared of the field actually introduces an additional scalar (under Lorentz transformations). In that sense, it is debatable whether this is actually just a vector field Lagrangian.

Of course,  we should also add to the action (\ref{eq:action}) all the matter terms  of
the standard model of particle physics. A general vector field action would  contain couplings of the vector field $A^\mu$ to this sector, as in bumblebee models. Such ``Lorentz-violating" interactions have been  thoroughly considered and constrained in the Standard Model Extension program initiated by Kosteleck\'y (see e.g. \cite{Colladay:1998fq,Kostelecky:2008ts}). In this article we assume that such couplings are negligible. Since interactions between the vector field and the matter sector would be mediated by gravitation, it is consistent to assume that they are small. For a review of modern tests of Lorentz invariance, see \cite{Mattingly:2005re}.

The validity of the effective theory proposed here requires that the background value of the vector field components satisfy ${|A^\mu| \ll \Lambda}$. This is the assumption that allows us to neglect terms in the Lagrangian such as, for instance,
\begin{equation}
	\mathcal{L}\supset \frac{\gamma}{\Lambda^2}A^\mu A^\nu \nabla_\mu A^\rho \nabla_\nu A_\rho.
\end{equation}
Because for a potential like $(\ref{eq:potential})$ we expect the field to be in the vicinity of ${A_\mu A^\mu=-M^2}$, this implies that $M^2\ll \Lambda^2$. This raises a mild fine-tuning  reminiscent of the hierarchy problem---why is the Higgs vev much smaller than the Planck scale. Up to an overall constant, a generic potential would be of the form
\begin{equation}
	V(A_\mu A^\mu)=2\lambda_1\Lambda^2 A_\mu A^\mu+\lambda_2  (A_\mu A^\mu)^2+\frac{\lambda_3}{\Lambda^2} (A_\mu A^\mu)^3+\cdots,
\end{equation}
with dimensionless coefficients $\lambda_i$ of order one. Comparing with equation (\ref{eq:potential}) we find that $\lambda=\lambda_2$ and $M^2=(\lambda_1/\lambda_2) \Lambda^2$. Thus, $M^2\ll \Lambda^2$ requires $\lambda_1\ll \lambda_2$.  This fine-tuning  does not have to be severe though. It suffices for instance that ${\lambda_1/\lambda_2\equiv M^2/\Lambda^2<10^{-2}}$ in order for our expression to be relatively accurate.  On a more practical side, the restriction $M<\Lambda$ is almost inescapable: If $M$ were of the order of $\Lambda$, we would have to include all powers of $A^\mu$  in our effective theory. Incidentally, a very small value of $\lambda$ can be easily justified if $\beta_{13}=\beta_4=0$. In that case, as we  discuss in Section \ref{sec:stability}, the theory has additional  symmetries in the limit $\lambda\to 0$.  For a review about  effective field theories see \cite{Burgess:2007pt}.

\comment{
\begin{figure}
\centering
\begin{picture}(200,46)
\includegraphics[height=8cm]{figura2.eps}
\put(-110,210){$A_{0}$}
\put(-20,120){$A_{i}$}
\put(-210,130){spacelike}
\put(-63,92){spacelike}
\put(-132,180){timelike}
\put(-132,35){timelike}
\end{picture}
\label{}       
\end{figure}
}

The action (\ref{eq:action}) is to some extent a generalization of previously considered  theories. Setting $V\equiv 0$ one recovers the class of vector-tensor theories studied in \cite{Will:1981cz}, while replacing the potential (\ref{eq:potential}) by a fixed-norm constraint leads to a subclass of the  Einstein-aether models considered in \cite{Jacobson:2000xp}, and the vector field theories studied by Carroll and Lim  \cite{Carroll:2004ai}.\footnote{Our parameters are related to those of \cite{Carroll:2004ai} by $\beta_{13}\equiv \beta_1+\beta_3$, $\beta\equiv \beta_1+\beta_2+\beta_3$, and by those of \cite{Will:1981cz} by $\beta_1\equiv 2\epsilon-\tau$, $\beta\equiv -\tau$, $\beta_{13}\equiv \eta-\tau$ and $\beta_4\equiv\omega$.} Although a fixed normed constraint seems to be difficult to justify from a particle physics perspective, general potentials $V(A_\mu A^\mu)$ have been argued to appear in theories in which gauge invariance is spontaneously broken \cite{Bjorken:1963vg,Kraus:2002sa, Ford:1989me}, and a fixed-norm constraint can be naively recovered from a potential of the form (\ref{eq:potential}), at least classically, by taking the limit $\lambda\to \infty$ (we shall make this connection more explicit below.)

We are ready now to explore some of the physical consequences of a vector field described by the action  (\ref{eq:action}). Varying  with respect to the field $A^\mu$ leads to the  equations of motion
\begin{equation}\label{eq:motionA}
	\beta_1 \nabla_\nu F^\nu{}_\mu+\beta \nabla_\mu (\nabla_\nu A^\nu)+
	\beta_{13} R_{\mu\nu} A^\nu+\beta_4 R A_\mu-\frac{dV}{dA^2}A_\mu=0.
\end{equation}
Because the covariant derivative and the metric commute, the equations we obtain by varying with respect to $A^\mu$ are equivalent to  the ones we get varying with respect to $A_\mu$.  In fact, because the transition from a vector to a form involves just a field redefinition, ${A^\mu\to A_\mu=g_{\mu\nu} A^\nu}$, assuming that the action (\ref{eq:action}) is a functional of a vector $A^\mu$ or a form $A_\mu$ leads to physically equivalent results both classically and at the quantum level, even though, curiously,  the corresponding expressions for the energy-momentum tensors $T_{\mu\nu}\equiv-(2/\sqrt{-g})(\delta S_A/\delta g^{\mu\nu})$ have different forms: If the action is taken to be a functional of a form $A_\mu$,  the energy-momentum tensor is
\begin{eqnarray}\label{eq:EMT}
	T^\mathrm{form}_{\mu\nu}&=& 
	2\beta_1  F_{\mu\rho}F_\nu{}^\rho
	+2\,\beta \left[\left(A^\rho\nabla_\rho(\nabla_\sigma A^\sigma)+
	(\nabla_\rho A^\rho)^2\right) g_{\mu\nu}-2A_{(\mu} \nabla_{\nu)}(\nabla_\rho A^\rho)\right]+\nonumber \\
	&+&\beta_{13}\left[2\nabla_\rho \nabla_{(\mu}(A_{\nu)}A^\rho)-\nabla_\rho\nabla_\sigma(A^\rho A^\sigma) g_{\mu\nu}-\Box(A_\mu A_\nu)-4 A^\rho R_{\rho(\mu}A_{\nu)}\right]+\nonumber \\
	&-&2\beta_4 \left[R_{\mu\nu}A_\rho A^\rho+R A_\mu A_\nu-\nabla_\mu\nabla_\nu(A_\rho A^\rho)+ \Box (A_\rho A^\rho)g_{\mu\nu}\right]+
	\nonumber \\
	&+&2\frac{dV}{dA^2}A_\mu A_\nu+ \mathcal{L} \, g_{\mu\nu},
\end{eqnarray}
while if the action is a functional of a vector $A^\mu$, the energy momentum tensor turns out to be
\begin{eqnarray}
T^{\text{vect}}_{\mu\nu}&=&
	T^\mathrm{form}_{\mu\nu}
	+4\beta_1  (\nabla_\rho F^\rho{}_{(\mu})A_{\nu)}
	+4\,\beta A_{(\mu} \nabla_{\nu)}(\nabla_\rho A^\rho) +4\beta_{13}A^\rho R_{\rho(\mu}A_{\nu)}+ 
	\nonumber \\
	&{}&+4\beta_4R A_\mu A_\nu-4\frac{dV}{dA^2}A_\mu A_\nu.
\end{eqnarray}
The additional terms in the energy-momentum tensor of a vector are proportional to the equation of motion (\ref{eq:motionA}), and thus vanish identically.  In the following, we work with the energy-momentum tensor of a form, equation (\ref{eq:EMT}). Note that $dA^2$ stands for $d(A_\mu A^\mu)$.

\section{Cosmological Solutions}\label{sec:cosmology}
On large scales, our universe seems to be well described by a (spatially flat)  Friedmann-Robertson-Walker (FRW) spacetime,
\begin{equation}\label{eq:FRW}
	ds^2=-dt^2+a^2(t) \, d\vec{x}\,^2.
\end{equation}
Its symmetries, homogeneity and isotropy, already constrain the possible configurations of a non-vanishing cosmic vector field quite strongly. Homogeneity (invariance under spatial translations) implies that a cosmic vector field must be spatially constant, while isotropy (invariance under spatial rotations) implies that its spatial components must vanish.\footnote{Except for some especial cases in which non-homogeneous field configurations are pure gauge so do not contribute to the energy-momentum tensor.} Hence, a cosmic vector field must be of the form
\begin{equation}\label{eq:A}
	A^\mu=(u(t),0,0,0).
\end{equation}

Because of homogeneity and isotropy, the  spatial components of  the vector field (\ref{eq:A}) satisfy the equations of motion (\ref{eq:motionA}), while the $\mu=0$ component of equation (\ref{eq:motionA}) dictates that $u$ has to obey
\begin{equation}\label{eq:motion}
	\beta \left(\ddot{u}+3H\dot{u}\right)+	\left[V'-3(\beta_{13}+4\beta_4)H^2+3(\beta-\beta_{13}-2\beta_4)\dot{H}\right]u=0,
\end{equation}
where  a dot denotes a derivative with respect to cosmic time $t$,  a prime  a derivative with respect to $A_\mu A^\mu$, and $H\equiv \dot{a}/a$ is Hubble's constant. It follows automatically from equation (\ref{eq:motion}) that if $\beta=0$, $u$ is non-dynamical.  Indeed,  in that case the ``kinetic" term of the vector is the one of electromagnetism, which does not contain any time derivatives of $A_0\equiv -u$. This is why there are no terms proportional to $\beta_1$ in the equation of motion for $u$. In aether models, the fixed norm constraint simply demands that the value of $u$ be constant. 

The  energy-momentum tensor (\ref{eq:EMT}) also explains why isotropy requires that the vector field have zero spatial components. Otherwise, say, the term proportional to $dV/dA^2$ would yield an energy-momentum tensor which would not be invariant under spatial rotations, and the FRW metric (\ref{eq:FRW}) would not be a solution of Einstein's equations. Certainly, if the vector field is subdominant, it is not strictly necessary that  $A^\mu$ satisfy equation (\ref{eq:A}) in order for equation (\ref{eq:FRW}) to be an \emph{approximate} solution of the Einstein equations. Ultimately, equation (\ref{eq:A}) should be justified dynamically, though we shall not explore this issue here. If the vector does point in the time direction, the energy density $\rho\equiv -T^0{}_0$ and pressure $p\, \delta^i_j \equiv T^i{}_j$ are given, from (\ref{eq:EMT}), by
\begin{eqnarray}
	\rho&=& \beta (2u\ddot{u}-\dot{u}^2) +6(\beta_{13}+2\beta_4)Hu \dot{u}+
	\left[2V'+6(\beta-\beta_{13}-2\beta_4)\dot{H}-9(\beta+2\beta_4)H^2\right]u^2+V, \nonumber \\
		  \\
	p&=&2(\beta-\beta_{13}-2\beta_4)(u\ddot{u}+\dot{u}^2)-\beta \dot{u}^2
	+(3\beta-2\beta_{13}-2\beta_4)\left[4 H u \dot{u}+(2\dot{H}+3H^2)u^2\right]-V.\nonumber \\	\label{eq:pressure}
\end{eqnarray}
All the non-diagonal components of the energy-momentum tensor vanish, as demanded by isotropy.  Because the Maxwell term does not contain $u(t)$, the energy density and pressure do not have factors proportional to $\beta_1$.  

It will be useful to work with a simplified form of the energy density. The equation of motion (\ref{eq:motion}) can be used to eliminate the term proportional to $\ddot{u}$ in $\rho$, leading to the simpler expression
\begin{equation}
	\rho=-\beta\dot{u}^2-6(\beta-\beta_{13}-2\beta_4)H u \dot{u}-3(3\beta-2\beta_{13}-2\beta_4)H^2 u^2+V. \label{eq:energy}
\end{equation}
Using the equation of motion again, it can be easily shown that the energy density (\ref{eq:energy}) satisfies the conservation equation $\dot{\rho}+3H(\rho+p)=0$, which is just an expression of the conservation law $\nabla_\mu T^\mu{}_\nu=0$. Note that $\rho$ may become negative for certain values of the parameters. In principle one should not be scared by the possible appearance of negative energies, as long as the Hamiltonian has a local minimum. We explore the classical and quantum stability of this class of theories in Section \ref{sec:stability}. The executive summary is: Unleashed aether theories are stable only if $\beta=0$ or $\beta_1=0$. Since  cosmological solutions with $\beta=0$ have no dynamics, we shall thus set $\beta_1=0$.\footnote{It is important to point out here that we are only considering \textit{classical} (coherent) excitations for the vector field. Theories with $\beta=0$ but $\beta_{1}\neq 0$ could also have cosmological relevance (as it happens for instance with the electromagnetic field) if they do not behave classically. (For instance, in the case of the Cosmic Microwave Background, what we have is a thermal distribution of photons).} In any case, this does not have any effect on our equations, because, as mentioned above, $\beta_1$ does not appear in the field equation of motion nor the energy-momentum tensor. For $\beta_{1}\neq 0\neq\beta$ the theory is unstable, at least in the ultraviolet. However, by an appropriate choice of the parameters we can get the dangerous modes out of the spectrum, recovering a sensible low energy effective field theory: the aether model.

\subsection{Solutions at early times}
We now turn our attention to the solutions of the vector field equation and its cosmological implications.  The equation of motion (\ref{eq:motion}) resembles that of a scalar field, with  additional contributions to its effective  mass   stemming from the expansion of the universe. We shall study the case in which this mass is dominated by the expansion, $H^2\gg V'$, which we generally expect to happen at early times, and also the case in which the mass is dominated by the potential term, $H^2\ll V',$ which we typically expect  to happen at late times. We take the expansion history as given, which in general requires the vector field to be subdominant. In some cases, if the equation of state of the vector field is compatible with the assumed expansion of the universe, our solutions also apply to periods in which the vector field is the dominant component in the universe.

In the limit $H^2\gg V'$, the equation of motion of the field takes the form
\begin{equation}
	\ddot{u}+3H\dot{u}+3\left(\frac{\beta-\beta_{13}-2\beta_4}{\beta} \dot{H} -\frac{\beta_{13}+4 \beta_4}{\beta} H^2\right)u=0.
\end{equation}
The behavior of the solution crucially depends on the size of the term in parenthesis, which is typically of order $H^2$. To find explicit solutions it is convenient and realistic to assume that the universe expands as a power law, $a\propto t^p$, which leads to the exact solution
\begin{eqnarray}\label{eq:ut}
	u(t)&=&C_1 \, t^{(1-3 p+f)/2}+C_2\,  t^{(1-3p-f)/2},  \quad \text{where}\\	
	f&=&\sqrt{(1+3p)^2+\frac{12p}{\beta}\left[\beta_{13}(p-1)+2\beta_4(2p-1)\right]},
\end{eqnarray} 
and $C_1$ and $C_2$ are two integration constants. To be more specific, let us  consider for example radiation domination, $p=1/2$. Keeping  the dominant mode only we get
\begin{equation}\label{eq:early evolution}
	u(t)\approx C_1 \, t^{(q-1)/4}, \quad \text{where}\quad q\equiv \sqrt{25-12\beta_{13}/\beta}.
\end{equation}
Thus, the field grows for $q>1$, decays for $q<1$, and oscillates logarithmically with decaying amplitude if $q$ is purely imaginary. Its equation of state  can then be calculated by substituting (\ref{eq:ut}) into equations (\ref{eq:energy}) and (\ref{eq:pressure}), though we shall not do so, as the resulting expression is not particularly illuminating.  A quite interesting dark energy model in which a massless vector  field evolves at early times following equation (\ref{eq:early evolution}) has been recently proposed in \cite{Jimenez:2008au}.

\subsection{Solutions at intermediate times} 

Generally we expect the condition $H^2\gg V'$ to break down at late times, either because $H$ steadily decreases, or because the field evolves to regions of larger potential slope  as the universe expands. In the limit $H^2\ll V'$, the equation  of motion (\ref{eq:motion}) resembles that of a minimally coupled scalar field,
\begin{equation}
	\ddot{u}+3H \dot{u}+\frac{V'}{\beta}u=0.
\end{equation}
Thus, even if the field is initially frozen, during a small time $\Delta t$  it  changes by an amount
\begin{equation}\label{eq:roll}
	\frac{\Delta u}{u} \approx -\frac{V'}{2\beta} \Delta t^2.
\end{equation}
It follows that during a Hubble time $\Delta t=H^{-1}$ the relative change of the field is of order $V'/(\beta H^2)$. If the latter is large, the field  rolls down until the potential slope is sufficiently small.

\subsection{Solutions at late times}\label{sec:extremum}

If the potential $V$ has the form (\ref{eq:potential}), the equations of motion in the absence of matter have  flat space and $\bar{A}^\mu=(M,0,0,0)$ as a solution.  We would speak of this configuration as the (classical) ``vacuum" of the theory. In an expanding universe however, $\bar{A}^\mu=(M,0,0,0)$ does not solve equation (\ref{eq:motion}), unless  spacetime is de Sitter and $\beta_{13}=-4\beta_4$. To study the properties of cosmological solutions around the vacuum, let us substitute into equation (\ref{eq:motion}) the ansatz $u=M+\delta u$, with $\delta u\ll  M$ and $V$ given by equation  (\ref{eq:potential}). Keeping only the leading terms we get
\begin{equation}\label{eq:linearized cosmo}
	 \beta [\delta\ddot{u}+3 H \delta\dot{u}] -4\lambda M^2 \delta u=
	 3M[(\beta_{13}+4\beta_4) H^2-(\beta-\beta_{13}-2\beta_4) \dot{H}].
\end{equation}
The configuration $u=M$ is stable if $-4\lambda M^2/ \beta>0$, which is what we assume in the following.  In this case, the equation describes a damped harmonic oscillator with a time-dependent external force. To solve  the equation we shall consider the limit  of large masses or times, that is $H/M\ll 1$ (we assume that $\dot{H}\approx H^2$.)  In this limit, one can think of the driving force as being constant, so an approximate particular solution of the inhomogeneous equation is
\begin{equation}\label{eq:adiabatic}
	\delta u_\mathrm{ad} =\frac{3}{4}\frac{(\beta-\beta_{13}-2\beta_4) \dot{H}-(\beta_{13}+4\beta_4) H^2}{\lambda M^2} M.
\end{equation} 
This is the leading correction to the solution $u=M$  in powers of the adiabaticity parameter $H/M$. Substituting equation (\ref{eq:adiabatic})  back into (\ref{eq:linearized cosmo}) one can indeed verify that corrections to this solution are suppressed by two additional powers of $H/M$. The derivatives of this adiabatic solution  are also  suppressed by powers of $H/M$. Not coincidentally, in the limit $H/M\ll 1$, the inequality $-4\lambda M^2/ \beta>0$  follows from one of the conditions for classical stability at low momenta that we derive in Section \ref{sec:stability aether}.

Having found a particular solution of the inhomogeneous equation, we can come back to the general solution of the homogeneous equation,
\begin{equation}\label{eq:oscillations}
	\delta u_\mathrm{osc}\approx \frac{B}{a^{3/2}} \cos\left(2 \sqrt{-\frac{\lambda}{\beta}} M t+\varphi \right),
\end{equation}
where $B$ is an arbitrary amplitude and $\varphi$ an arbitrary phase. During a Hubble time the field performs $ \sqrt{-\lambda/\beta} M/H\gg 1$ oscillations, while its amplitude remains essentially constant. 

We have seen that the full solution of equation (\ref{eq:linearized cosmo}) has the form
\begin{equation}\label{eq:full}
	u=u_\mathrm{ad}+\delta u_\mathrm{osc},
\end{equation}
where $\delta u_\mathrm{osc}$ is given by equation (\ref{eq:oscillations}), and, to lowest order, $u_\mathrm{ad}\approx M$. In order to find the energy density and pressure associated with the full solution (\ref{eq:full}), we substitute it into equations (\ref{eq:energy}) and (\ref{eq:pressure}). Because we are interested on cosmological timescales, we consider time averages of $\rho$ and $p$ over Hubble times. Among the terms quadratic in $u_\mathrm{ad}$, those without derivatives dominate. Terms that involve mixed terms $u_\mathrm{ad} \, \delta u_\mathrm{osc}$ average to zero, while the averages of terms quadratic in $\delta u_\mathrm{osc}$ give non-zero contributions proportional to the square of the amplitude $B$.  We can thus split the contributions to the energy density in an adiabatic component, and an oscillatory one.

To leading order,  the energy density and pressure of the adiabatic solution are,
\begin{eqnarray}
	\rho_\mathrm{ad}&=&-3 (3\beta-2\beta_{13}-2\beta_4)  H^2 M^2 ,\\
	p_\mathrm{ad}&=& (3\beta-2\beta_{13}-2\beta_4) (3 H^2+2 \dot{H}) M^2.
\end{eqnarray}
The subleading terms are suppressed at least by a factor of order $H^2/M^2$, and, hence, their  contributions to $\rho$ and $p$, of order $H^4$, are negligible.  Because the Friedmann equations in the presence of matter can be cast as 
\begin{equation}
	3 H^2= 8 \pi G_\mathrm{c} \rho_m \quad \text{and} \quad -(3H^2+2\dot{H})=8\pi G_\mathrm{c} p_m,
\end{equation}
an observer in such a universe trying to determine $G_\mathrm{c}$ through the expansion of the universe would find a renormalized value of Newton's constant,
 \begin{equation}\label{eq:cosmic G}
 	G_\mathrm{c}=\frac{G}{1+8\pi G M^2 (3\beta-2\beta_{13}-2\beta_4)}.
 \end{equation}
As we discuss in Section \ref{sec:stability aether}, stability requires that $3\beta-2\beta_{13}>0$.

Exactly the same renormalization  occurs in aether models \cite{Carroll:2004ai}. We show in Section \ref{sec:PPN} that for generic parameter values $G_\mathrm{c}$ also differs from the gravitational constant measured in ``local" solar system measurements $\mathrm{G}_\mathrm{N}$.  Following the analysis of \cite{Carroll:2004ai}, one can then derive a not too stringent order of magnitude limit on the scale $M$ based on nucleosynthesis constraints,   
\begin{equation}\label{eq:looseBBNbound}
	8 \pi G M^2 \lesssim 1.
\end{equation}
Since our analysis assumes that $M<\Lambda$, and presumably $\Lambda$ is smaller than the scale of quantum gravity $M_G\equiv (8\pi G)^{-1/2}$, this is also a necessary condition for the validity of our approximations. Hence, we typically expect $G_c\approx G$.  Nevertheless, for some parameter choices $G$ and $G_c$ can differ significantly. If $8\pi G M^2 (3\beta-2\beta_{13}-2\beta_4)$ is fine-tuned to be close to $-1$, the Planck scale one would derive from $G_c$ would be much lower than the actual quantum gravity scale $M_G$. On the other hand, if $8\pi GM^2 \gg 1$,  the actual scale of quantum gravity would be much lower  than the one guessed from the value of $G_c$. Of course, this would require $\Lambda\gg M_G$.

The density and pressure of the oscillating solutions are dominated by those terms with the maximum number of derivatives acting on $u$. Thus, substituting equation (\ref{eq:oscillations}) into (\ref{eq:energy}) and (\ref{eq:pressure}) and averaging over time we find that the oscillating vector field behaves like non-relativistic matter,
\begin{equation}
	\rho_\mathrm{osc}\approx 4\lambda M^2 \frac{B^2}{a^3} \quad \text{and} \text \quad p_\mathrm{osc}\approx 0.
\end{equation}
The  effective pressure is consistent with the scale factor dependence of the energy density.  Since we assume that the vector field only couples to gravity, there is no additional decay channel through which the vector field could dissipate its energy (if $\beta_{1}=0$ the spectrum of the model consists of a single scalar.) Thus, an oscillating vector field is a potential dark matter candidate, somewhat similar to the axion, though we would have to study how structure develops to confirm that the vector actually mimics  dark matter. These oscillations are absent in aether models, because the constraint on the vector field norm only allows the lowest order adiabatic component. Note that in  the limit in which $\lambda\to \infty$, for fixed $M$, the adiabatic approximation $u_\mathrm{ad}\approx M$ becomes exact, and the energy density of the  oscillating solution diverges, unless $B=0$.

The energy density of the oscillating field depends on the  value of the amplitude $B$, and therefore,  constraints on the scale $M$ based on this solution depend on initial conditions.  Nevertheless, based on the adiabatic solution (\ref{eq:adiabatic}), which does not depend on initial conditions, it appears reasonable to assume that  when $H\approx M$, the amplitude of the oscillations was $(B/a_M)\approx M/\lambda$. Under this assumption, demanding that the energy density of the oscillating field during nucleosynthesis be no more than the total energy density we arrive at the bound
\begin{equation}
	M\lesssim  ( \lambda^2 M_P^3\, T^2_\mathrm{nucl})^{1/5}\sim  \lambda^{2/5}\, 10^{10}\, \textrm{GeV},
\end{equation}
where $T_\mathrm{nucl}\sim \textrm{MeV}$ is the temperature during nucleosynthesis, and $M_P\approx 10^{18}\, \mathrm{GeV}$ is the reduced Planck mass.  For $\lambda$ of order one this limit improves the bound (\ref{eq:looseBBNbound}) by eight orders of magnitude.

\section{Newtonian Limit}\label{sec:PPN}

The presence of the  vector field not only affects gravity on cosmological scales, but also locally.  In a massless vector field theory for instance, the vector field determines the gravitational constant that enters the Newtonian limit of the theory, and the post-Newtonian parameters of the theory generically differ from those of general relativity \cite{Will:1981cz}. In a massive theory, one would expect the effects of the vector field to be negligible at sufficiently large scales, but  it turns out that, even on large scales, the vector field leaves its imprint on the effective gravitational constant that appears in the Poisson equation. In this section we study some aspects of the Newtonian and Post-Newtonian limits of the unleashed aether. A general discussion of the Post-Newtonian limit in the gravitational sector of Kosteleck\'y's standard model extension  \cite{Colladay:1998fq,Kostelecky:2008ts}  can be found in \cite{Bailey:2006fd}.

We would like to determine the gravitational field created by a non-relativistic and static source sitting in otherwise empty space. For a potential like (\ref{eq:potential}) the equations of motion in the absence of matter have Minkowski spacetime as a solution,
\begin{equation}\label{eq:background solution}
	ds^2=-dt^2+d\vec{x}^2, \quad \bar{A}^\mu=(M,0,0,0).
\end{equation}
We expect the presence of a sufficiently small and static matter source, with $T_{\mu\nu}=\rho_m \delta_{\mu 0} \delta_{\nu 0}$, to modify this solution slightly, so we can use first order perturbation theory to calculate its gravitational field. Because the energy density $\rho_m$ is a scalar under spatial rotations, in this limit it suffices to restrict our attention to the scalar sector. Let us hence perturb
\begin{equation}
	ds^2=-(1+2\Phi) dt^2+(1-2\Psi)d\vec{x}\,^2, \quad A^\mu=\bar{A}^\mu+(u,\partial_i s),
\end{equation}
and solve the equations of motion to first order in the perturbations.  If we were interested, say, in gravitational radiation we would have to study the vector and tensor sectors as well. 

In Fourier space the $\mu=0$ and $\mu=i$ linearized field equations  (\ref{eq:motionA}) are
\begin{eqnarray}
	(\beta_1 k^2-4\lambda M^2)(u+M \Phi)&=&(\beta_{13}-\beta_1-2\beta_4)  k^2 M\Phi
	-4 \beta_4 k^2 M (\Psi-\Phi), \label{eq:linearized} \\
	\beta k^2 k_i s&=&0 \label{eq:linearized i}.
\end{eqnarray}
Because we are interested in solutions that vanish at infinity, it follows from  equation (\ref{eq:linearized i}) that $s=0$. As we discuss in Section \ref{sec:stability aether}, stability in this background requires either $\beta_1=0$, or  $\beta_1>0$ and $\lambda<0$. Hence, the relative signs of the operator multiplying $u+M \Phi$ in equation (\ref{eq:linearized}) are correct.  The same equation can be combined with  the linearized $i-j$ Einstein equations
\begin{equation}\label{eq:linearized ij}
	(\Phi-\Psi)(1-16\pi G M^2 \beta_4)=32\pi G \beta_4 M(u+M\Phi),
\end{equation}
to solve for the value of Eddington parameter $\gamma$, implicitly defined by the relation $\Psi=\gamma \Phi$. If $\beta_4=0$, $\gamma=1$ follows directly from (\ref{eq:linearized ij}), but in the general case the solution is not that simple,
\begin{equation}\label{eq:gamma scale}
	\gamma=\frac{(\beta_1 k^2 -4\lambda M^2)(1-16\pi G M^2 \beta_4)-32\pi GM^2k^2\beta_4(\beta_{13}-\beta_1+2\beta_4)}{(\beta_1 k^2 -4\lambda M^2)(1-16\pi G M^2 \beta_4)-128\pi G M^2 k^2\beta_4^2}.
\end{equation}

Inspection of equation (\ref{eq:gamma scale})  reveals that the Eddington parameter is scale-dependent, and that the scale in the problem is set by  the squared mass of the scalar,  $-\lambda M^2$. Therefore,  for coefficients $\beta_i$ of order one there are two relevant regimes: $k^2 \ll |\lambda M^2|$ (large scales), and $k^2 \gg |\lambda M^2|$ (small scales). The value of $\gamma$ in both regimes is
\begin{equation}\label{eq:gamma}
\gamma=
	\left\{
	\begin{array}{cr}
	\displaystyle 1 & \text{large scales},\\
	\displaystyle  \frac{\beta_1+16\pi G M^2 \beta_4(\beta_1-2\beta_{13} -4\beta_4)}{\beta_1-16\pi G M^2\beta_4(\beta_1+8\beta_4)} &\text{small scales}.
	\end{array} \right.
\end{equation} 
On short scales the value of $\gamma$ agrees  with the one calculated for massless vectors in \cite{Will:1981cz}, while on large scales it approaches the value in aether models \cite{Eling:2003rd,Foster:2005dk} and general relativity $\gamma=1$ (the case $\beta_1=\beta_4=0$ requires special consideration, and is analyzed below.) At present, the best measurement of $\gamma$ comes from Doppler tracking of the Cassini spacecraft \cite{Bertotti:2003rm}, which gives $\gamma-1=(2.1\pm 2.3)\times 10^{-5}$. In the limit $\lambda\to \infty$, all lengths fall in the long wavelength regime. 

It is also instructive to study how $u$ is related to the potential $\Phi$. Combining again equations (\ref{eq:linearized}) and (\ref{eq:linearized ij}) we can solve for $u$ in terms of $\Phi$. To keep the expression manageable, let us write down the solution in terms of the linear perturbation of a quantity proportional to the perturbation of the  squared norm $A_\mu A^\mu$,
\begin{equation}\label{eq:delta AA}
	u+M\Phi=\frac{ (\beta_1-\beta_{13}+2\beta_4) \left(16 \pi G M^2 \beta_4-1\right)}
	{k^2[\beta_1-16 \pi G M^2 \beta_4 (\beta_1+8 \beta_4)] -4 \lambda M^2(1-16\pi G  M^2\beta4)}
	 M k^2 \Phi.
\end{equation}
In the massive case the vector field perturbation $u$ is still non-negligible even far away from the matter source. In fact, at long distances,  $u$ approaches $-M\Phi$, which typically decays as $1/r$, where $r$ is the distance to the source.  This is quite different from a massive scalar-tensor theory, in which to first order the scalar field perturbation $\delta\varphi$ satisfies the equation  $(k^2+m^2)\delta \varphi \propto \rho_m$, and thus vanishes exponentially far away from matter, $\delta \phi\propto \exp(-m r)/r$.  Clearly, the origin of the difference is  that $u+M\Phi$ plays the role of the short-ranged scalar $\delta\varphi$. 

Once we know how $\Psi$ and $u$ are related to $\Phi$ we can proceed to solve for the gravitational potential $\Phi$.  The linearized 00 Einstein equation is
\begin{equation}\label{eq:linearized 00}
	-k^2 \Psi=4\pi G 
	\left\{\left[(2\beta_{13}+4\beta_4)k^2 -8\lambda M^2\right]M(u+M\Phi)-2\beta_{13}k^2 M^2\Phi+4\beta_4k^2 M^2(\Psi-\Phi)+\rho_m\right\}.
\end{equation}
Inserting $\Psi=\gamma \Phi$ and  equation (\ref{eq:delta AA}) into (\ref{eq:linearized 00}) leads to the Poisson equation $\Delta \Phi =4\pi G_\mathrm{eff} \rho_m$. However, the effective gravitational constant $G_\mathrm{eff}$ is actually not constant,  but scale dependent,
\begin{equation}\label{eq:Newtonian G}
	\frac{G_\mathrm{eff}}{G}=
	\left\{
	\begin{array}{cl}
	[1-8 \pi G M^2(\beta_1+2\beta_4)]^{-1} & \text{large scales}, \\
	\displaystyle \frac{\beta_1}{\beta_1 \gamma -8\pi G M^2 [2\beta_1(\beta_{13}-\beta_4(\gamma-2))-(\beta_{13}+2\beta_4)(\beta_{13}+2\beta_4(1-2\gamma))]} & \text{small scales}.
	\end{array} \right.
\end{equation}
The vector field again renormalizes $G$ on large and small scales, but by a different amount than in the cosmological case (\ref{eq:cosmic G}), even on large scales. This disagreement in the local and cosmological values of $G$ points to a violation of Birkhoff's theorem in these theories, and also  is what led to the nucleosynthesis constraint discussed in Section \ref{sec:cosmology}. Note that the change in the gravitational constant caused by the term proportional  to $\beta_4$  is the one that follows from replacing $A^2\to -M^2$ in the action (\ref{eq:action}), and is the same both for the cosmological and  Newtonian gravitational constants.  If we set $\beta_4=0$,  our value of $G_\mathrm{eff}$ on large scales agrees with the one found for aether models in \cite{Carroll:2004ai}. Note that the value of $\beta$ does not have any impact on the Newtonian limit of the theory. 

As an illustration of these results, let us consider a massive vector  with $\beta_4=0$. It follows automatically from equation (\ref{eq:gamma}) that $\gamma=1$, so in this case the theory agrees with the current measurements of $\gamma$ both on short and large scales. The effective Newton constant takes the simpler form
\begin{equation}
	\frac{G_\mathrm{eff}}{G}=
	\left\{
	\begin{array}{cc}
	[1-8\pi G M^2 \beta_1]^{-1}& \text{large scales},\\
	\displaystyle [1-8\pi G M^2 \beta_{13}(2-\beta_{13}/\beta_1)]^{-1} & \text{small scales}.
	\end{array} \right.
\end{equation}
This equation  is remarkable because it says that the strength of  gravity changes with scale. It is hence tempting to speculate a connection with dark matter or even dark energy.  For instance,  if $\beta_1/\beta_{13}>0$ gravity is stronger at large scales than at small scales, so if  the vector had a Compton wavelength of a few kiloparsecs this could eventually explain the flatness of galactic rotation curves without dark matter.   Recall that if $\beta_1\neq 0$, stability imposes $\beta_1>0$ and $\beta_{13}>0$. 

Such a scale-dependent behavior of the gravitational coupling  does not occur in massless vector field theories or aether models, though it does happen in theories with more complicated Lagrangians \cite{Zlosnik:2006zu}.    In connection with modified Newtonian dynamics \cite{Milgrom:1983ca}, a specific model that involves a fixed-norm vector has been proposed by Bekenstein in \cite{Bekenstein:2004ne}. Ideas about degravitation on large  scales have been explored in \cite{Dvali:2007kt} and references therein.

\begin{center}
 * \quad * \quad *
\end{center}

Although our expressions (\ref{eq:gamma}) and (\ref{eq:Newtonian G}) still hold when  $\beta_1=0$ or $\beta_4=0$, they become ill-defined when both $\beta_1$ and $\beta_4$ vanish. This case however is  particularly simple. It follows directly from equation (\ref{eq:linearized ij}) that $\gamma=1$, while the remaining linearized equations lead to
\begin{equation}
	\frac{G_\mathrm{eff}}{G}=\frac{4\lambda M_G^2}{4\lambda M_G^2-\beta_{13}^2 k^2}.
\end{equation}
Therefore, on short scales the theory does not have a Newtonian limit (the gravitational potential satisfies the equation $k^4\Phi\propto \rho$), while if the field is massive, $G_\mathrm{eff}\approx G$ on long scales.

\section{Ghosts and Instabilities}\label{sec:stability}

The strongest constraints on the parameters in the action (\ref{eq:action}) are set by the requirement that the theory be classically and quantum-mechanically stable.  In this somewhat technical section we proceed to analyze the conditions that classical and quantum-mechanical stability place  on the parameters of the action (\ref{eq:action}). Analogous conditions for  vector fields of constrained norm have been derived in \cite{Lim:2004js,Carroll:2008em}, while a somewhat similar analysis of massless vector fields of arbitrary norm has been carried out in \cite{Jimenez:2008sq}. Instabilities in models of inflation supported by vector fields have been recently reported in \cite{Himmetoglu:2008zp,Himmetoglu:2008hx,Himmetoglu:2009qi}. Our stability analysis closely resembles the one for massive gravity  in \cite{Dubovsky:2004sg}

In order to analyze the stability of unleashed aether models, we consider vector field  fluctuations $b_\mu$ around the background value $\bar{A}_\mu$ of the  field, 
\begin{equation}\label{eq:fluctuations}
	A_\mu=\bar{A}_\mu+b_\mu.
\end{equation}
Because we are mostly interested in cosmological solutions we assume that the background vector  $\bar{A}^\mu$ has the form (\ref{eq:A}), and leave the case of a field that points along a spatial direction to Appendix \ref{sec:appendix}. 
We neglect metric perturbations, which relies on the assumption that $\Lambda\ll M_G$.  In conformal time coordinates, the metric and background field are hence given by
\begin{equation}\label{eq:background}
	ds^2=a^2(\eta)(-d\eta^2+d\vec{x}\,^2), \quad
\bar{A}^\mu=a^{-1}\cdot (\bar{A},0,0,0).
\end{equation}
Note that $\bar{A}$ plays the role of $u$ in Section \ref{sec:cosmology}, and is hence dynamical. 

To obtain the Lagrangian of the fluctuations $b^\mu$  we insert  the expansion (\ref{eq:fluctuations}) and the background (\ref{eq:background}) into the action (\ref{eq:action}). Expanding to quadratic order  in the fluctuations we arrive at
\begin{eqnarray}\label{eq:action a}	\mathcal{L}_b=&-&\beta_{1}\partial_{\mu}b_{\nu}\partial^{\mu}b^{\nu}-(\beta-\beta_1)(\partial_{\mu}b^{\mu})^{2}
	-4\beta(\mathcal{H}^2 b_0^2 -\mathcal{H}b_0\partial_\mu b^\mu) +\nonumber \\
	{}&+&\beta_{13}R_{\mu\nu}b^\mu b^\nu+a^2\beta_4R \,b_\mu b^\mu-a^2 V' b_\mu b^\mu-2a^4V'' \cdot (\bar{A}^\mu b_\mu)^2 ,
\end{eqnarray}
where indices are raised with the  Minkowski metric $\eta^{\mu\nu}$,  $\mathcal{H}=d\log a/d \eta$, and a prime denotes a derivative with respect to $A_\mu A^\mu$.  For arbitrary parameters  $\beta_1$ and $\beta$, the four components of the vector field are dynamical, because the determinant of the Hessian matrix 
\begin{equation}\label{eq:Hessian}
	\frac{\partial^2 \mathcal{L}_b}{\partial (db_\mu/d\eta) \partial(db_\nu/d\eta)}
\end{equation}
is non-zero \cite{Henneaux:1992ig}.  However, as we shall see below, for $\beta=0$ or $\beta_1=0$ the determinant of the Hessian vanishes, signaling that some of the vector field components are constrained. This is what happens for instance for the gauge invariant Maxwell theory ($\beta=0$). We thus study the degenerate cases $\beta=0$ and $\beta_1=0$ separately.   In aether models, the fixed norm constrain requires $b^0=0$ \cite{Lim:2004js}, and thus eliminates one of the four dynamical fields. Note by the way that if we expand the potential of equation (\ref{eq:potential}) to quartic order in the perturbations we obtain an additional term
\begin{equation}
	\mathcal{L}_b\supset \lambda (b_\mu b^\mu)^2.
\end{equation}
Thus, for $\lambda>1$, the theory is strongly coupled. 

\subsection{Stability in the vector sector}

The background value of the vector field determines the set of unbroken spacetime symmetries. Because $\bar{A}_\mu$ points along the time direction, the action (\ref{eq:action a}) is symmetric under spatial translations and rotations. It is  convenient to decompose the fields in irreducible representations of this unbroken symmetry group. Expanding as usual
\begin{equation}\label{eq:action fluctuations}
b_0\equiv -u \quad \text{and} \quad b_i=\partial_i r+v_i, \quad \text{with} \quad \partial_i v^i=0,
\end{equation}
decouples the two scalar modes  $u$ and $r$ from the two vector modes in $\vec{v}$. The Lagrangian of the vector sector is then given by
\begin{equation}\label{eq:vector L}
	\mathcal{L}_v=-\beta_1  \partial_\mu \vec{v}\cdot \partial^\mu \vec{v}- a^2 m_v^2 \, \vec{v}\cdot \vec{v},
\end{equation}
where
\begin{equation}\label{eq:mv}
 m_v^2\equiv  V'-3(\beta_{13}+4\beta_4)H^2-(\beta_{13}+6\beta_4)\dot{H}.
\end{equation}
The factor of $a^2$ in the mass term of equation (\ref{eq:vector L}), as well as the contributions to the mass proportional to the curvature explicitly break time translation invariance. Therefore, the quantity one would naively identify as the frequency of the modes in the vector sector
\begin{equation}\label{eq:omega v}
	\omega_v^2=\vec{p}\,^2+\frac{a^2}{\beta_1}m_v^2
\end{equation}
also depends on time (in this regard one can consider
$a^2 m^2_v/\beta_1$ as the ``mass'' of these excitations.) We can nevertheless quantize the theory under the approximation that the frequency is constant if  the relative change in $\omega$ during the characteristic time $\Delta \eta \approx \omega^{-1}$ is small \cite{Mukhanov:2007zz}, 
\begin{equation}\label{eq:adiabaticity}
	\frac{1}{\omega^2}\frac{d\omega}{d\eta}\ll 1.
\end{equation}
In the short wavelength regime, $\vec{p}\, ^2/a^2 \gg m_v^2$, this adiabaticity condition is automatically satisfied, but in the long wavelength regime $\vec{p}\, ^2/a^2 \ll m_v^2$, it imposes restrictions on the magnitude of  $V'$ and its time derivatives. For instance,  if $V'\ll H^2$, the mass $m_v$ is of the order of $H^2$, and hence $\omega^{-2}_v (d\omega_v/d\eta)$ is of order one.  Our analysis in this section  exclusively applies in the adiabatic regime (\ref{eq:adiabaticity}). In this regime, we can  set all explicit factors of  $a$ to one. 

Returning now to equation (\ref{eq:vector L}), and expressing the vector in terms of two orthogonal transverse polarizations immediately reveals that the vector sector is ghost free for ${\beta_1\geq 0}$, and free of classical instabilities at low momenta ($\vec{p}\,^2\ll |m_v^2|$) for $m_v^2/\beta_1\geq 0.$ These conditions are summarized in Table \ref{table:vector stability}.  We should also  point out that tachyonic, long-wavelength instabilities in a theory may be allowed, as long as they are not too severe. Their severity depends on the magnitude of the tachyonic ``mass" $\omega^2\equiv -m_t^2$, which determines the rate of growth of the mode, proportional to $\exp(m_t \, \eta)$. Since we mostly assume that all masses are much larger than the Hubble scale, the presence of a tachyon would imply instabilities on timescales much shorter than the Hubble time, which is likely to be problematic. On the other hand, tachyonic masses of the order of $H$ (or smaller) should not be much of a problem. Note that if $\beta_1=0$, the vector sector is not dynamical. 

\begin{table}
\begin{tabular}{|c|c|c|}\hline
{\bf Vector} & Low $p$ & High $p$ \\ \hline 
 Classical & $\beta_1 m_v^2\geq 0$ & \checkmark \\ \hline
Quantum & $\beta_1 \geq 0$ & $\beta_1 \geq 0$ \\ \hline
\end{tabular}
\caption{Stability conditions in the vector sector. The checkmark means that the condition is automatically satisfied. The mass $m_v$ is defined in equation (\ref{eq:mv}).}\label{table:vector stability}
\end{table}

\subsection{Stability in the scalar sector}

To proceed further, we expand the scalar fields in appropriately normalized Fourier modes, 
\begin{equation}\label{eq:expansion}
	u=\int \frac{d^4p}{(2\pi)^{4}} \, u(p) \exp(i p_\mu x^\mu), \quad
	r= \int \frac{d^4p}{(2\pi)^{4}} \, \frac{r(p)}{|\vec{p}\,|} \exp(i p_\mu x^\mu).
\end{equation} 
Since the fields are real, it follows that $u^*(p)=u(-p)$ and $r^*(p)=r(-p)$. Substituting the expansions (\ref{eq:expansion}) in (\ref{eq:action a}) and assuming as above that all explicitly time-dependent quantities are constant we arrive at the action (in Fourier space)
\begin{equation}\label{eq:timelike L}
\mathcal{S}_s=-\int \frac{d^4p}{(2\pi)^{4}}
\left(\begin{array}{l}u \\ r\end{array}\right)^\dag D \left(\begin{array}{l}u\\ r\end{array} \right),
\end{equation}
where the matrix $D$ is given by 
\begin{equation}
D=\left(\begin{array}{lr}
\beta \, \omega^2-\beta_1\vec{p}\,{}^2+ m_u^2 & -i\,\omega\, |\vec{p}\,| (\beta-\beta_1)-2\beta |\vec{p}\,| H  \\
i\,\omega\, |\vec{p}\,| (\beta-\beta_1)-2\beta  |\vec{p}\,|H & -\beta_1\,\omega^2+\beta\,\vec{p}\,{}^2+m_v^2  
\end{array}
\right)
\end{equation}
and we have also defined
\begin{equation}\label{eq:mu}
	m_u^2\equiv 2V''\bar{A}^2-V'+(2\beta+3\beta_{13}+12\beta_4)H^2-(2\beta-3\beta_{13}-6\beta_4)\dot{H}.
\end{equation}
In spite of our notation, the mass parameter $m_u^2$ is not the mass of the field $u$, because the Lagrangian is not diagonal in that field. Note that in an expanding universe, the terms proportional to $R_{\mu\nu}$ and $R$ in equation (\ref{eq:action}) also contribute to  $m_u$ and $m_v$. These contributions are typically negligible, unless the derivatives of the potential are sufficiently small.

The inverse of the  matrix $D$ is the field propagator. Thus,  in order to find the propagating modes we just have to find the values of $\omega^2$ at which  its eigenvalues have poles, or, equivalently, the values of $\omega^2$ at which the eigenvalues of $D$ have zeros. The determinant of $D$ is
\begin{equation}\label{eq:determinant}
	\text{det}\, D=-\beta\beta_1 \omega^4+(2\beta\beta_1 \vec{p}\,^2-\beta_1 m_u^2+\beta  m_v^2)\omega^2-(\beta\beta_1\vec{p\,}^4-\beta m_u^2 \, \vec{p}\,^2+\beta_1 m_v^2\,  \vec{p}\,^2+4\beta^2 H^2 \vec{p}\,^2 -m_u^2 m_v^2).
\end{equation}
Requiring that  $\det D$ vanish we thus arrive at the frequencies of the \emph{two} propagating scalar modes
\begin{eqnarray}
	\omega^2&=&\vec{p}\,^2+\frac{m_v^2}{2\beta_1}-\frac{m_u^2}{2\beta}\pm \Delta, \quad \text{where} \label{eq:omega}\\
	\Delta&\equiv& \sqrt{\left[\left(\frac{1}{\beta_1}-\frac{1}{\beta}\right) (m_u^2+m_v^2)-4\frac{\beta}{\beta_1} H^2\right]\vec{p}\,^2 +\frac{1}{4}\left(\frac{m_u^2}{\beta}+\frac{m_v^2}{\beta_1}\right)^2}. \label{eq:delta}
\end{eqnarray}
Observe that some of the factors of $\omega^2$ in the determinant (\ref{eq:determinant}) are multiplied by the mass terms $m_v^2$ and $m_u^2$. Since factors of $\omega$ arise from time derivatives, and because the mass terms contain derivatives of the potential, we may say that the potential is also part of the  vector field kinetic terms.

The theory is free of classical instabilities (exponentially growing modes) if the frequency  $\omega$ in equation (\ref{eq:omega}) is real. 
In the limit of  low momenta, in which $\vec{p}\,^2$ is negligible, the two frequencies approach\footnote{We are assuming that $(m_u^2/\beta+m_v^2/\beta_1)<0$. Otherwise the values of $\omega_+$ and $\omega_-$ become interchanged.}
\begin{eqnarray}\label{eq:omega low p}
	\omega^2_+ \to &-&\frac{m_u^2}{\beta}
	+\left[1-\frac{(\beta-\beta_1)(m_u^2+m_v^2)-4 \beta^2H^2}{\beta_1m_u^2+\beta m_v^2}\right]\vec{p}\,^2
	+\mathcal{O}\left(\vec{p}\,^4\right),\\ 
	\omega^2_-\to &{}&\frac{m_v^2}{\beta_1}
	+\left[1+\frac{(\beta-\beta_1)(m_u^2+m_v^2)-4 \beta^2H^2}{\beta_1m_u^2+\beta m_v^2}\right]\vec{p}\,^2+\mathcal{O}\left(\vec{p}\,^4\right),
\end{eqnarray}
in which, for later convenience, we have included terms that are typically subdominant at low momenta. In the limit of high momenta,  in which mass terms can be neglected, the frequencies become
\begin{equation}\label{eq:omega high p}
	\omega_\pm^2\to \vec{p}\,^2\pm |\vec{p}|
	\sqrt{\left(\frac{1}{\beta_1}-\frac{1}{\beta}\right) (m_u^2+m_v^2)-4\frac{\beta}{\beta_1}H^2}+\mathcal{O}\left(|\vec{p}\,|^0\right).
\end{equation}
Demanding that the mode frequencies be real at low and high momenta we arrive at the conditions on Table \ref{table:scalar stability}, which if satisfied actually imply that the system is classically stable for all momenta (this follows from a detailed analysis of equations (\ref{eq:omega}) and (\ref{eq:delta}).)  Inspection of equations (\ref{eq:omega}) and (\ref{eq:delta}) shows that the regimes of high and low momentum not only depend on the masses of the  fields, but also on the values of $\beta$ and $\beta_1$.  Note that in the high-momentum limit, the leading term in the frequency gives $\omega_\pm\approx \vec{p}\,^2$. We consider the subleading contributions, proportional to $|\vec{p}\,|$, because if the latter are imaginary there are instabilities on arbitrarily short timescales.

The residues at the poles of the eigenvalues of $D^{-1}$ determine whether the propagating modes are ghosts. If these residues have negative signs, the corresponding modes have negative energies. In the presence of ghosts, the vacuum can then decay into positive energy quanta and negative energy ghosts, leading to a quantum-mechanically unstable theory. How severe this instability is depends on the way the ghosts couple to positive energy quanta, and on the volume of phase space available for the decay \cite{Carroll:2003st,Cline:2003gs}. The coupling of ghosts to positive energy quanta is essentially determined by the coupling constants of the theory and the  magnitude of the residues, while the phase space  available for decay typically depends on the cut-off  up to which the theory is valid. Choosing any of these parameters appropriately might tame  ghost instabilities, although we shall not explore this possibility here.  In this context, it is worthwhile pointing out however that with Lorentz-invariance spontaneously broken, the cutoff of the theory involves \emph{spatial} momenta rather than Lorentz-covariant four-momenta. To illustrate this let us consider one of the many higher-dimensional operators that we neglected in (\ref{eq:action}),
\begin{equation}\label{eq:corrections}
	\frac{\delta}{\Lambda^6}A^\mu A^\nu A^\rho A^\sigma (\nabla_\mu \nabla_\nu A^\tau) (\nabla_\rho\nabla_\sigma A_\tau),\label{eq:higher dimensional}
\end{equation}
where  $\delta$ is a dimensionless parameter of order one.  Expanding (\ref{eq:corrections}) to quadratic order in the fluctuations $b_\mu$  in Minkowski space we obtain
\begin{equation}\label{eq:correction}
	\mathcal{L}_s\supset \delta\frac{\bar{A}^4 \omega^4}{\Lambda^6}u^2. 
\end{equation}
Hence, this term would make an effective contribution to the matrix $D$ which would become comparable to the lowest order term $\beta\omega^2 u^2$ if
 \begin{equation}
 	\omega^2\approx \frac{\Lambda^6}{\bar{A}\,^4}.
\end{equation}
As we have seen, at low momenta the dispersion has the form $\omega^2=m^2$, where, by definition, $m$ is the mass of the field, while at high momenta the dispersion relation typically approaches $\omega^2=c_s^2\, \vec{p}\,^2$, where $c_s^2$ is a coefficient of order one. Thus,  our effective description is certainly not valid for  masses and spatial momenta beyond $\Lambda^3/\bar{A}^2$. In fact, higher-dimensional operators analogous to (\ref{eq:corrections}), of the form
\begin{equation}
	\frac{1}{\Lambda^{4n-2}}A^{\mu_1}\cdots A^{\mu_n} 
	A^{\rho_1}\cdots A^{\rho_n}
	(\nabla_{\mu_1}\cdots  \nabla_{\mu_m} A^\tau)  
	(\nabla_{\rho_1}\cdots  \nabla_{\rho_m} A_\tau), 
\end{equation}
point to a breakdown scale of the order $\Lambda^2/\bar{A}$. Note that in the limit  $\bar{A}\to 0$, the corrections due to the new term goes to zero. In that limit Lorentz invariance is restored, and large spatial momenta can be made small by a Lorentz transformation.\footnote{A similar phenomenon was described in the context of a scalar field theory in an inflating universe in \cite{ArmendarizPicon:2008yv}. There, the scale of Lorentz symmetry breaking is set by the Hubble constant, which plays the role of $\bar{A}$ here.}

 If $\lambda_+(\omega^2)$ and $\lambda_-(\omega^2)$ are the two eigenvalues of the matrix $D$,  and the former have zeros at, respectively, $\omega_+$ and $\omega_-$, the residues at the poles of $D^{-1}$ are\footnote{In some cases, one of the eigenvalues may have multiple zeros. In that case, evaluation of equation (\ref{eq:Z}) at those zeroes gives the corresponding residues.}  
\begin{equation}\label{eq:Z}
	Z_\pm=\lim_{\omega^2 \to\omega_{\pm}^2} \frac{(-\omega^2+\omega_{\pm}^2)}{\lambda_{\pm}(\omega)}\quad \Rightarrow \quad \frac{1}{Z_\pm}=-\frac{d\lambda_{\pm}}{d\omega^2}\Bigg|_{\omega^2_\pm}.
\end{equation}
The theory is free of ghosts if both residues are positive. Since the two eigenvalues of the matrix $D$ are 
\begin{equation}
	\lambda_{1, 2}= \frac{\text{tr}\, D \pm \sqrt{\text{tr}^2  D-4\, \text{det}\, D}}{2}
\end{equation}
the condition for the absence of ghosts is then 
\begin{equation}\label{eq:no ghosts}
	\frac{1}{Z_{\pm}}=-\left(\frac{1}{\text{tr}\, D}{\frac{d\,  \text{det}\, D}{d\omega^2}}\right)\Bigg|_{\omega^2_{\pm}} >0.
\end{equation}

The trace of the matrix $D$ at $\omega_\pm$ is
\begin{equation}
	\text{tr} D\big|_{\omega^2_{\pm}}=2(\beta-\beta_1) \vec{p}\,^2+\frac{\beta+\beta_1}{2}\left(\frac{m_u^2}{\beta}+\frac{m_v^2}{\beta_1}\right)\pm (\beta-\beta_1) \Delta,
\end{equation}
whereas  the derivative of the determinant is 
\begin{equation}\label{eq:derivative D}
\frac{d\,  \text{det}\, D}{d\omega^2}\Bigg|_{\omega^2_{\pm}}=\mp 2\beta\beta_1 \Delta.
\end{equation}
From the previous equations it immediately follows that the theory always contains a high-momentum ghost. Indeed, in the limit of high momenta, the trace approximately equals the same value both at $\omega_+$ and $\omega_-$, while the value of $d \det D/d\omega^2$ changes sign. Therefore, the residues cannot be positive for both frequencies. More precisely, in the limit of high-momentum the residues approach
\begin{equation}\label{eq:Z generic}
	\frac{1}{Z_\pm}\to \pm \frac{\sqrt{(\beta_1^{-1}-\beta^{-1})(m_v^2+m_u^2)-4\beta\beta_1^{-1} H^2}}{(\beta_1^{-1}-\beta^{-1})|\vec{p}\,|},
\end{equation}
which implies that one linear combination of fields is a ghost, while the remaining orthogonal combination is stable (classical stability forces the term inside the square root to be positive.)  Because at low momenta the residues approach
\begin{equation}\label{eq:Z low p}
	\frac{1}{Z_+}\to -\beta,\quad \frac{1}{Z_-}\to \beta_1,
\end{equation}
 it is possible to avoid the existence of low-momentum ghosts for appropriate choices of $\beta$ and $\beta_1$. Classical and quantum stability conditions are summarized in Table \ref{table:scalar stability}.

Note by the way that there is a ghost at high momenta  for $\beta=\beta_1$, even though equation (\ref{eq:Z generic}) breaks down for those values. In this case, the trace of $D$ has always the same value, irrespectively of $\vec{p}$.  If $H$ or any of the two scalar masses  is non-zero, $\Delta$ is non-zero, and, as before, equation (\ref{eq:no ghosts}) cannot be satisfied. This is immediately apparent in the Lagrangian (\ref{eq:action a}). For $\beta=\beta_1$, the Lagrangian   is already diagonal in the fields $b^\mu$, and because of the indefinite signature of the Minkowski metric, at least one of the fields has to be a ghost.

\begin{table}
\begin{tabular}{|c|c|c|}\hline
 {\bf Scalar} ($\beta_1\neq 0, \, \beta\neq 0$) & Low $p$ & High $p$ \\ \hline 
 Classical & $ \beta_1\cdot  m_v^2>0,\, \beta\cdot m_u^2<0$ &   $(\beta_1^{-1}-\beta^{-1})(m_u^2+m_v^2)-4\beta\beta_1^{-1} H^2>0$ \\
 \hline
Quantum & $ \beta_1>0, \beta<0$ & \XSolid \\ \hline
\end{tabular}
\caption{Stability conditions in the scalar sector  for  $\beta_1\neq 0$ and $\beta\neq 0$. The cross means that the condition cannot be satisfied. The parameters $m_v$ and $m_u$ are defined, respectively, in equations (\ref{eq:mv}) and (\ref{eq:mu}).}\label{table:scalar stability}
\end{table}

\subsubsection{Massless case}

It is important to realize that the high-momentum limit is different from the strict massless case
\begin{equation}
	H^2=m_u^2=m_v^2=0.
\end{equation}
Indeed, in the massive case there are always two propagating modes in the scalar sector at high momenta, with frequencies that approach the relativistic dispersion relations ${\omega^2=\vec{p}\,^2(1\pm \mathcal{O}(m/|\vec{p}\,|))}$.  In contrast, in the massless case, $\Delta$ in equation (\ref{eq:delta}) vanishes, and equation (\ref{eq:derivative D}) implies that there is a double pole at $\omega^2=\vec{p}\,^2$ instead of two single poles at $\omega^2=\vec{p}\,^2$. Thus, in the massless case there is a scalar with higher time derivatives plus the two degrees of freedom of the vector sector. Lagrangians with higher time derivatives are usually quantized  using a procedure developed by Ostrogradski, and are typically  unstable (see \cite{Woodard:2006nt} for a review.)

The same conclusion follows by substituting the Lorentz-covariant decomposition ${b_\mu=\partial_\mu \phi+v_\mu}$, with $\partial_\mu v^\mu=0$ into the action
\begin{equation}\label{eq:S_M}
	S_M=\int d^{4}x \,\left[ -\frac{\beta_{1}}{2}F_{\mu\nu}F^{\mu\nu}-\beta(\nabla_{\mu}A^{\mu})^{2}-m^2 \,A_\mu A^\mu\right]. 
\end{equation}
This yields a Lagrangian which describes massive electrodynamics in ``Lorentz gauge'' and a scalar $\phi$ whose kinetic term contains  higher order derivatives,
\begin{equation}\label{eq:Lorentz L}
	\mathcal{L}_M=-\frac{\beta_1}{2} (\partial_\mu v_\nu-\partial_\nu v_\mu)(\partial^\mu v^\nu-\partial^\nu v^\mu)-\beta (\Box \phi)^2-m^2 \partial_\mu \phi \partial^\mu \phi -m^2 v^\mu v_\mu.
\end{equation}
In this form, it is manifest that the Lagrangian not only describes a vector field, but also a scalar field (under Lorentz transformations.)
By inspecting  the scalar kinetic term in equation (\ref{eq:Lorentz L}) one would naively conclude that  in the massive case there  are two scalar poles, at $p^2=-m^2/\beta$ and $p^2=0$, but this is illusory. Because $\partial_\mu v^\mu=0$, the propagator of the original vector field $A_\mu$  is
\begin{equation}\label{eq:propagator A}
	\Delta_A=\Delta_v+p_\mu p_\nu \Delta_\phi=
	\frac{1}{\beta_1 p^2+m^2}\left(\eta_{\mu\nu}-\frac{p_\mu p_\nu}{p^2}\right)+
\frac{1}{\beta \, p^2+m^2}\frac{p_\mu p_\nu}{p^2}.
\end{equation}
The propagator of $v$ is just that of electrodynamics in $R_\xi$ gauge \cite{Weinberg:Rxi}, with $\xi\to 0$ and an added mass term. It can be obtained by inverting the kinetic term in equation (\ref{eq:S_M}) and letting $\beta$, which plays the role of $1/\xi$, tend to infinity. This propagator is manifestly transverse (projects onto the subspace perpendicular to $p_\mu$), while the contribution from $\phi$ to $\Delta_A$ is clearly longitudinal (projects onto the subspace parallel to $p_\mu$.) In the form (\ref{eq:propagator A}) it is now clear that there are no poles at $p^2=0$. Instead, there is a pole at $p^2=-m^2/\beta$ in the longitudinal sector, and three poles at $p^2=-m^2/\beta_1$ in the transverse sector. These correspond to the same frequencies that we found in our previous analysis, as the reader can readily verify by setting $H=0$, $m_v^2=m^2$ and $m_u^2=-m^2$ in equations (\ref{eq:omega}) and (\ref{eq:delta}). Note that we could have also obtained (\ref{eq:propagator A}) by directly inverting the differential operator in equation (\ref{eq:Lorentz L}). 

Let us finally discuss $m^2=0$, as we intended to do. In this case, the poles at $p^2=0$ that canceled in the massive case remain, and give rise to factors proportional to $1/p^4$. There is then a double pole at $p^2=0$, and two single poles at $p^2=0$. These considerations also illustrate that a free ($V\equiv 0$)  vector field in flat space can behave  very differently from a free  vector field in an expanding universe if $\beta, \beta_{13}$ or $\beta_4$ are different from zero.

\subsubsection{Degenerate Limits}
Our conclusions do not apply to the two degenerate limits in which equation (\ref{eq:omega}) breaks down namely, for $\beta=0$ and $\beta_1=0$. In both cases, the determinant of the Hessian matrix (\ref{eq:Hessian}) vanishes, or equivalently, the determinant of $D$, equation (\ref{eq:determinant}), is not quartic in $\omega$, but quadratic. The Lagrangian therefore describes only one scalar propagating mode, instead of two.  

When $\beta=0$, the ``kinetic" part of the Lagrangian is that of the electromagnetism and the Proca Lagrangian.  The determinant of the matrix $D$ only has one zero at
\begin{equation}\label{eq:b0 omega}
	\omega_0^2=\frac{m_v^2}{\beta_1}-\frac{m_v^2}{m_u^2}\,\vec{p}\,^2,
\end{equation}
and the residue at this pole is 
\begin{equation}\label{eq:b0 r}
	Z_0=\frac{1}{\beta_1}+\frac{m_v^2-m_u^2}{m_u^4}\, \vec{p}\,^2.
\end{equation} 
Thus, equations (\ref{eq:b0 omega}) and (\ref{eq:b0 r}) result in the stability conditions in Table \ref{table:b0b1 stability}. Note that some of these conditions are the same as in the vector sector, Table \ref{table:vector stability}, and that if the theory is stable both in the low and high momentum limits, it will be stable for all momenta.  If $m_u^2=0$, the equation $\text{det}\, D=0$ has no solution  and the theory does not have any propagating scalar degrees of freedom. From the point of view of flat space, equation (\ref{eq:propagator A}), setting $\beta=0$ eliminates the longitudinal polarization associated with the scalar $\phi$. In the massless case, one of the the remaining transverse polarizations disappears, leaving just the two modes of the vector sector. In this case, the theory has an additional symmetry, gauge invariance. 

As illustration, let us consider the familiar example of the Proca Lagrangian, in which $\beta_1=1/2$, $\beta=\beta_{13}=\beta_4=0$ and $V=m^2 A_\mu A^\mu/2$. For a vanishing background field this  implies $m_v^2=m^2/2$ and $m_u^2=-m^2/2$. It follows from Table \ref{table:b0b1 stability}, that the Proca field is stable for $m^2>0$ and both classically and quantum-mechanically unstable for $m^2<0$. Again, the massless limit is not continuous: for $m^2=0$ there are no propagating scalars.  Massive versions of electromagnetism with mass terms that violate Lorentz invariance ($m_u^2\neq -m_v^2$) have been studied in \cite{Dvali:2005nt,Gabadadze:2004iv} .
\begin{table}
\parbox{8cm}
{
\begin{tabular}{|c|c|c|}\hline
{\bf Scalar} ($\beta= 0$)& Low $p$ & High $p$ \\ \hline 
Classical & $\beta_1 m_v^2>0$ & $m_v^2m_u^2<0$ \\ \hline
Quantum & $\beta_1> 0$ & $ m_v^2-m_u^2 >0$ \\ \hline
\end{tabular}
}
\begin{minipage}{8cm}
{
\begin{tabular}{|c|c|c|}\hline
{\bf Scalar} ($\beta_1= 0$)& Low $p$ & High $p$ \\ \hline 
Classical & $ \beta m_u^2<0$ & $m_v^2(m_u^2-4\beta H^2)<0$ \\ \hline
Quantum & $\beta<0$ & $ m_u^2-m_v^2-4\beta H^2<0$ \\ \hline
\end{tabular}
}
\end{minipage}
\caption{Stability conditions in the scalar sector  for timelike background fields in the degenerate cases $\beta_1=0$ or $\beta= 0$. If the theory is stable both in the low and high momentum limits, it follows from equations (\ref{eq:b0 omega}),  (\ref{eq:b0 r}), (\ref{eq:b1 omega}) and (\ref{eq:b1 r})  that it is actually stable for all momenta. The parameters $m_v$ and $m_u$ are defined, respectively, in equations (\ref{eq:mv}) and (\ref{eq:mu}).}\label{table:b0b1 stability}
\end{table}

If $\beta_1=0$ the determinant of $D$ vanishes at
\begin{equation}\label{eq:b1 omega}	
	\omega_0^2=-\frac{m_u^2}{\beta}-\frac{m_u^2-4\beta H^2}{m_v^2}\vec{p}\,^2,
\end{equation}
while the residues at the poles are
\begin{equation}\label{eq:b1 r}
	Z_0=-\frac{1}{\beta}-\frac{m_v^2-m_u^2+4\beta H^2}{m_v^4}\vec{p}\,^2.
\end{equation} 
Combining both (\ref{eq:b1 omega}) and (\ref{eq:b1 r}) we arrive at the stability conditions in Table \ref{table:b0b1 stability}. If these conditions are satisfied, the theory is stable for all momenta. Once again, if $m_v^2=0$  the inverse of $D$ has no poles, and the theory does not contain any dynamical scalar (and, since $\beta_1=0$, there is no dynamical vector either.) In a Lorentz-covariant language, setting $\beta_1=0$ eliminates the three transverse  poles  in  the propagator of equation (\ref{eq:propagator A}). In the massless case, the longitudinal polarization also disappears, leaving a theory without dynamics. In fact, in this limit the theory has an additional symmetry $b_\mu\to b_\mu+\varepsilon_\mu{}^{\nu\rho\sigma}\partial_\nu c_{\rho\sigma}$, where $c_{\rho\sigma}$ is an antisymmetric tensor (a two-form), which can be used to gauge all the components of $b_\mu$ away. 

\subsection{Stability around the minimum of the potential}
\label{sec:stability aether}

As an important application of this analysis, let us consider stability in the case where the vector field sits at the minimum of a potential given by equation (\ref{eq:potential}). As discussed in Subsection \ref{sec:extremum} this is generically possible only in flat space.  Substituting $H\equiv0$ and equation (\ref{eq:potential}) into equations (\ref{eq:mv}) and (\ref{eq:mu}) we find 
\begin{equation}
	m_v^2=0\quad \text{and} \quad m_u^2=4\lambda M^2.
\end{equation}
Let us consider the case $\beta_1\neq 0$ and $\beta\neq 0$ first. The spectrum then contains two vector and two scalar degrees of freedom, with dispersion relations at low momenta given, respectively, by equations (\ref{eq:omega v}) and (\ref{eq:omega low p}),
\begin{equation}\label{eq:spectrum}
\omega^2_{v_1}=\vec{p}\,^2, \quad \omega^2_{v_2}=\vec{p}\,^2,\quad
\omega^2_{s_+}\approx-\frac{4\lambda M^2}{\beta}, \quad \omega^2_{s_-}\approx\frac{\beta}{\beta_1}\vec{p}\,^2.
\end{equation}
The three ``massless" excitations $v_1, v_2$ and $s_-$ can thus be interpreted as the Nambu-Goldstone bosons associated with the spontaneous breaking of the three generators of Lorentz boosts.  At low momenta, the corresponding residues are, from equations (\ref{eq:vector L}) and (\ref{eq:Z low p}),
\begin{equation}\label{eq:residues}
Z_{v_1}=1/\beta_1, \quad Z_{v_2}=1/\beta_1,\quad
Z_{s_+}=-1/\beta, \quad Z_{s_-}=1/\beta_1.
\end{equation}
In consequence, at low momenta the theory is either classically or quantum-mechanically stable, but not both. The absence of ghosts requires $\beta_1>0$ and $\beta<0$, which then implies that $s_-$ is classically unstable. Conversely,  classical stability demands $\lambda/\beta<0$ and $\beta/\beta_1>0$, which then implies that either $v_1, v_2$ and $s_-$, or $s_+$, are ghosts. The only way  to escape this conclusion is to choose a well-behaved set of massless modes ($\beta>0$ and $\beta_1>0$) and push the massive mode---either a tachyon or a ghost---out of the domain of validity of the effective theory.  Indeed, as mentioned above, if the mass of $s_+$ is of the order $\Lambda^2/M$, we cannot trust our analysis anyway, so the presence of a tachyon or ghost in the spectrum is not necessarily fatal. In principle, this pathology could be solved by the ultraviolet completion of the theory. Note that these instabilities cannot be avoided by lowering the mass scale $M$ and ``squeezing" the low-momentum regime out of the spectrum of the theory; at high momenta,  $\vec{p}\,^2\gg 4\lambda M^2$, according to Table \ref{table:scalar stability} and equation (\ref{eq:Z generic}), either $s_+$ or $s_-$ are ghost-like too.

Formally, we can decouple the massive mode $s_+$ by sending $|\lambda|$ to infinity, which is the limit in which we expect to recover aether models.  In fact, the low momentum spectrum (\ref{eq:spectrum}) and residues (\ref{eq:residues})  of the massless modes $v_1, v_2$ and $s_-$ agree with those of aether theories \cite{Lim:2004js}. In that sense, unleashed aether models can be regarded as a high-momentum completions of the former, which, as we have argued, are inconsistent for generic values of $\beta$ and $\beta_1$, unless $\lambda$ is larger than $(\Lambda/M)^4$. This is not just a fine-tuning problem. Because we expect $\Lambda/M\gg 1$, it also means that the theory is strongly coupled. The only way in which strong coupling issues could be circumvented (without avoiding fine-tuning) in a sensible low energy effective theory is by considering $0<\beta\lesssim(M/\Lambda)^{4}$. In any case,  from this  perspective it is also easy to understand the agreement on large scales of the Newtonian limits of both constrained and unleashed aether models. At low momenta, both are essentially the same theory.
  
These instabilities can be averted by setting $\beta_1=0$ or $\beta=0$. Since no non-trivial cosmological solutions with $\beta=0$ exist, let us hence consider $\beta_1=0$ next.   In this case the vector sector is non-dynamical, and because $m_v^2=0$, there are no dynamical scalars in the theory either. This is however a degenerate limit. In an expanding universe, the field is slightly displaced from the minimum, as discussed in Section \ref{sec:cosmology}.  Assuming that $H^2\ll 4\lambda M^2$ and inserting equation (\ref{eq:adiabatic}) into equation (\ref{eq:mv}) we find, to leading order, 
\begin{equation}\label{eq:mv minimum}
m_v^2=-(2\beta_{13}-3\beta)\dot{H}.
\end{equation}
 Hence, $m_v^2$ is of order $H^2$ and generally different from zero. The spectrum of the theory then consists of a single scalar $s_0$ with dispersion relation and residues  given by, respectively,
\begin{equation}
	\omega_0^2\approx -\frac{4\lambda M^2}{\beta}\left(1-\frac{\beta}{2\beta_{13}-3\beta} \frac{\vec{p}\,^2}{\dot{H}}\right) \quad \text{and} \quad
	\quad Z_0\approx -\frac{1}{\beta}+\frac{4\lambda M^2 \vec{p}\,^2}{(2\beta_{13}-3\beta)^2 \dot{H}^2}.
\end{equation}
Thus, the scalar is stable at low momenta if $\beta<0$ and $\lambda>0$. Because $\dot{H}<0$ whenever the null energy condition\footnote{$\rho\geq 0$ and $\rho+p\geq 0$.} is satisfied, the scalar is stable at high momenta if in addition $2\beta_{13}-3\beta<0$. Comparing with the general case above, equation (\ref{eq:spectrum}), we see that at low momenta the effect of setting $\beta_1=0$  is to remove the (classically unstable) scalar Goldstone mode $s_-$ from the spectrum. The remaining scalar is massive, and thus is absent from the low-momentum description provided by aether theories. In fact, aether models with $\beta_1=0$ do not have any propagating degrees of freedom.

The case $\beta=0$ is somewhat similar. Using equation (\ref{eq:mv minimum}) we find  that the dispersion relation and residues of the only scalar mode are 
\begin{equation}
\omega_0^2\approx-2\beta_{13}\dot{H}\left(\frac{1}{\beta_1}-\frac{\vec{p}\,^2}{4\lambda M^2}\right),
\quad \text{and} \quad
Z_0\approx \frac{1}{\beta_1}- \frac{\vec{p}\,^2}{4\lambda M^2}
\end{equation}
Thus, using $\dot{H}<0$ again, stability at low momenta requires $\beta_1>0$ and $\beta_{13}>0,$ while stability at high momenta then imposes $\lambda<0$.

\subsection{Relation between energy conditions, instabilities and superluminality}

A violation of the standard energy conditions is often associated with  superluminal propagation and instabilities in a theory \cite{Carroll:2003st,Dubovsky:2005xd}. In k-field models for example, the absence of ghosts implies the null energy condition $\rho+p\geq 0$, and also places at the same time restrictions on the value of the speed of sound (see for instance \cite{ArmendarizPicon:2005nz}). It is therefore instructive to check whether these relations also extend to  unleashed aether models.

From equation (\ref{eq:omega}) one could  immediately calculate, say, the group velocity of the perturbations,
\begin{equation}
	v_g=\frac{d\omega}{dp}.
\end{equation}
For instance, it follows from equation (\ref{eq:spectrum}) that the propagation of $s_-$ is subluminal for $\beta/\beta_1<1$, and superluminal for $\beta/\beta_1>1$. As discussed in Section \ref{sec:stability aether}, the stability of that mode (in our restricted sense)  only requires $\beta>0$ and $\beta_1>0$. Similarly, from equation (\ref{eq:omega high p}), at high momenta the dispersion relations of the scalar modes approach the relativistic expressions $\omega_{\pm}=p$, whilst our stability analysis indicates that in that limit there is always a ghost in the spectrum. Thus, subluminal propagation seems to be independent from stability considerations (in this respect, the unleashed aether is  analogous to k-fields.) The requirement of subluminal  propagation is sometimes used to constrain the parameters in this type of theories, but we believe that there is nothing inconsistent with faster than light propagation in this background. The constant vector field defines a preferred frame and explicitly breaks Lorentz invariance.  Even if a signal propagates superluminally, it will move forward in our preferred time coordinate, it will  not be possible to construct closed timelike curves, and no conflict with causality will arise.  For different viewpoints on the issue see e.g. \cite{Adams:2006sv,Bruneton:2006gf,Babichev:2007dw,Dubovsky:2008bd}. Readers concerned about  superluminal propagation should impose appropriate conditions on the parameters of the theory.  For instance, the additional constraint $\beta/\beta_1\leq 1$ implies that the mode $s_-$ propagates subluminally (at low momenta).  Analogous conditions in the gravitational wave sector have been discussed in \cite{Lim:2004js}. Note that signal propagation in the vector sector is never superluminal.

In the limit in which the expansion of the universe is negligible the energy density (\ref{eq:energy}) is
\begin{equation}
	\rho=-\beta\dot{u}^2+V.
\end{equation}
In the same limit, the sum of pressure and energy density is
\begin{equation}	\rho+p=2\left(\frac{\beta_{13}+2\beta_4}{\beta}-1\right)V'u^2-2(\beta_{13}+2\beta_4)\dot{u}^2.
\end{equation}
Therefore, the energy density is positive for $\beta<0$ and $V>0$, while $\rho+p$ is positive if $\beta_{13}+2\beta_4<0$  and  $V'(\beta_{13}+2\beta_4-\beta)/\beta>0$. Examination of Tables \ref{table:vector stability} to \ref{table:b0b1 stability} shows that these conditions bear a marginal resemblance with the conditions for the absence of ghosts in the limit of low momentum, which is the one that applies for an homogeneous field. In particular, the absence of ghosts implies $\beta<0$, which is the condition for the ``kinetic'' part of the energy density to be positive.  Otherwise, at least in this particular case, it appears that the weak and null energy conditions have little to do with the presence or absence of ghosts, at least in this class of models.

\section{Summary and Conclusions}
At sufficiently long timescales and large distances, any generally covariant theory that contains a real vector field coupled to gravity, and is invariant under $A_\mu\to -A_\mu$, should be  described by an action of the form (\ref{eq:action}).  Massless vector-tensor theories \cite{Will:1981cz} and aether models \cite{Jacobson:2000xp} can be regarded as particular limits of  this ``unleashed aether" class.  We have identified some of their common features, and pointed out in what other aspects they differ significantly. 

Stability imposes severe conditions on  the  parameters of the action (\ref{eq:action}). The spectrum  of the theory generally consists of a massive excitation and three massless fields, which can be interpreted as the Nambu-Goldstone bosons associated with the spontaneous symmetry breaking of invariance under boosts. At least one of these four fields is always a tachyon or a ghost. For appropriate choices of parameters however, it is possible to decouple the dangerous modes. Setting $\beta_1=0$ eliminates all the  transverse components (in a Lorentz covariant way) of the vector. And setting $\beta=0$ eliminates just the longitudinal polarization, which  is dynamical only for a massive vector or in an expanding universe. Both choices result in a well-behaved theory. Non-trivial cosmological solutions exist only if $\beta_1=0$, that is, if the  kinetic term of the vector field is proportional to the squared divergence of the vector.  We have also extended our stability analysis to models in which the vector field points along a spatial direction (see the Appendix \ref{sec:appendix}) and found that they also suffer generically from ghosts.

There is yet another alternative to eliminate the unstable fields from the theory. For appropriate choices of parameters, the only unstable field in the theory is massive. If its mass  is sufficiently heavy, its excitations  are beyond  the regime of validity of our low-energy description and can be ignored. The remaining three fields, the  Nambu-Goldstone bosons, are the only fields present in the spectrum of aether models. In that sense the aether  can be regarded as a low-momentum description of unleashed aether models.  

Since the symmetries of the universe only encompass spatial translations and rotations, it is possible for  a vector field to have a non-vanishing expectation value along its time direction. In unleashed aether models such a non-vanishing cosmic vector field leads to a host of interesting gravitational phenomena. In the  Newtonian limit of the theory  Newton's constant is typically scale-dependent, and for some parameter values the Eddington parameter $\gamma$ differs from its value in general relativity. On cosmological scales, the vector field leaves an imprint on the expansion of the universe, even in the vicinity of a potential minimum. The field renormalizes Newton's constant, like in aether models, but it also typically oscillates around it. Under  relatively mild assumptions, these oscillations  strongly constrain the value of the field at the minimum of the potential.

Perhaps the most important conclusion is that self-consistent and phenomenologically acceptable modifications of gravity are difficult to come by. In this article we have added a massive vector field of unknown origin  to  the gravitational sector, and we have employed a low-energy effective approach to study the implications of this addition. We have seen that general vector field theories have a very rich gravitational phenomenology, but that self-consistency and agreement with basic cosmological and solar system observations impose severe constraints on this class of models.  Nevertheless,  a large volume of parameter space is still allowed by our requirements, and  many interesting phenomenological aspects remain to be investigated. Unleashed aether models can wander far away.

\begin{acknowledgements}
We thank Riccardo Penco for useful comments on an earlier version of this manuscript, and Quentin G. Bailey for pointing out to  us a few missing references.  ADT acknowledges the kind hospitality of the members of the Physics Department at Syracuse University, where this work was initiated. The work of CAP was supported in part by the US National Science Foundation under grant PHY-0604760. The work of ADT was supported by a UNAM postdoctoral fellowship  and funds from the Basque Government GICO7/51-IT-221-07.
 \end{acknowledgements}

\begin{appendix}
\section{Stability of Spacelike Vectors}\label{sec:appendix}

The stability analysis in Section \ref{sec:stability} assumes that the vector field points in the time direction, as required by  isotropy.  But even if the field is aligned along a spatial direction it is possible in some cases to  restore the isotropy of spacetime by either considering appropriate configurations of multiple vector fields \cite{ArmendarizPicon:2004pm,Golovnev:2008cf}, or by simply assuming that the vector is subdominant. Our stability analysis can also be extended to this type of configurations, for which the background value of  $A^\mu$ is spacelike. 

If the vector field points along a spatial direction, we can always pick our coordinates so that $\bar{A}^\mu$ points in the $z$ direction,
\begin{equation}\label{eq:spacelike A}
	\bar{A}^\mu=a^{-1}(0,0,0,\bar{A}).
\end{equation}
In order to determine whether such a configuration is stable, we consider again the fluctuations around the background value of the field, $A_\mu=\bar{A}_\mu+b_\mu$.  But instead of expanding the fluctuations as in (\ref{eq:action fluctuations}) we decompose the perturbations in irreducible representations of $SO(2)$, the group of rotations that leaves $\bar{A}^\mu$ in equation (\ref{eq:spacelike A}) invariant,
\begin{equation}\label{eq:spacelike expansion}
	b_0\equiv -\tilde{u}, \quad b_\alpha\equiv\partial_\alpha \tilde{r}+\tilde{v}_\alpha \,\, (\text{with}
	\, \partial_\alpha \tilde{v}^\alpha=0), \quad b_z\equiv \tilde{t},
\end{equation}
where $\alpha$ runs over the transverse spatial coordinates, $\alpha=x,y$. Thus, under $SO(2)$, $\tilde{u}$, $\tilde{r}$ and $\tilde{t}$ are scalars, while $\tilde{v}$ is a vector.

Substituting the expansion (\ref{eq:spacelike expansion}) into equation (\ref{eq:action a}) we obtain the Lagrangian for the perturbations. The Lagrangian for the transverse vector modes $\tilde{v}^\alpha$ takes the simple form 
\begin{equation}
	\mathcal{L}_{\tilde{v}}=-\beta_1 \partial_\mu \tilde{v}_\alpha \partial^\mu \tilde{v}^\alpha-m_v^2\tilde{v}_\alpha \tilde{v}^\alpha, 
\end{equation}
where $m_v^2$ is given by equation (\ref{eq:mv}) (recall that we assume that $a^2 m_v^2$ is adiabatically constant, and thus set $a= 1$.) Thus, the absence of instabilities leads again to the conditions in Table \ref{table:vector stability}. For spacelike background fields the scalar sector in this case is slightly more complicated, because there are three coupled scalar fields, instead of two. In Fourier space, their action is
\begin{eqnarray}\label{eq:spacelike L}
	\mathcal{S}_{\tilde{s}}=-\frac{1}{2}\int \frac{d^4p}{(2\pi)^{4}}&& \Big\{\left[\beta \, \omega^2 -\beta_1 \vec{p}\,^2+\tilde{m}_u^2\right] \tilde{u}^*\tilde{u}+
	\left[-\beta_1 \omega^2 \vec{p}_{\perp}{}^2
	+\beta_1 \vec{p}\,^2 p_{\perp}^2+(\beta-\beta_1)\vec{p}\,^4_{\perp}+p_{\perp}^2 m_v^2\right] \tilde{r}^*\tilde{r} + \nonumber \\
	&&+\left[-\beta_1 \omega^2	+\beta_1\vec{p}\,^2+(\beta-\beta_1)p_{\parallel}^2+\tilde{m}_t^2\right] \tilde{t}^*\tilde{t}
	-4\beta H \vec{p}_{\perp}\!^2 \tilde{r}^*\tilde{u}+\\
	&&-4i\beta H p_\parallel\,\tilde{t}^*\tilde{u}
	-2(\beta-\beta_1)\left[i\omega \vec{p}_{\perp}\!^2\tilde{r}^* \tilde{u}-\omega p_{\parallel} \tilde{t}^*\tilde{u}
	-i \vec{p}_\perp\!^2 p_{\parallel}\, \tilde{t}^*\tilde{r}\right]+h.c.
	\Big\}, \nonumber
\end{eqnarray}
where we have defined
\begin{eqnarray}	\tilde{m}_u^2&=&-V'+(2\beta+3\beta_{13}+12\beta_4)H^2-(2\beta-3\beta_{13}-6\beta_4)\dot{H}, \label{eq:mu tilde} \\ 	\tilde{m}_t^2&=&V'+2V''\bar{A}^2-3(\beta_{13}+4\beta_4)H^2-(\beta_{13}+6\beta_4)\dot{H}, \label{eq:mt tilde}
\end{eqnarray}
and $p_{\parallel}=p_z$, $\vec{p}_{\perp}=(p_x,p_y,0)$. Note that $\tilde{m}_u^2$ differs from the $m_u^2$ in equation (\ref{eq:mu}).  For later convenience, we have not normalized $s$ canonically.  

A general analysis of the stability of this system would be rather cumbersome, because we would have to diagonalize the $3\times3$ matrix.  For our purposes however it suffices to focus on particular values of the momentum $\vec{p}$, for which some of the scalars decouple from the others.  Specifically, setting $\vec{p}_{\perp}=0$ in equation (\ref{eq:spacelike L}) we arrive at the action 
\begin{eqnarray}
\mathcal{S}_{\tilde{s}}=-\int \frac{d^4p}{(2\pi)^{4}} &\Big\{&\left[\beta \, \omega^2 -\beta_1  p_\parallel^2+\tilde{m}_u^2\right] \tilde{u}^*\tilde{u}
	+\left[-\beta_1 \omega^2 +\beta p_{\parallel}^2+\tilde{m}_t^2\right] \tilde{t}^*\tilde{t}+\nonumber \\
	{}&+&2i\beta H p_\parallel\left[\tilde{u}^* \tilde{t}- \tilde{t}^* \tilde{u}\right]
	+(\beta-\beta_1) \omega p_{\parallel}\left[\tilde{u}^* \tilde{t}+\tilde{t}^* \tilde{u}\right] \Big\}.
\end{eqnarray}
This has the same form as  equation (\ref{eq:timelike L}). The field $\tilde{r}$ is non-dynamical and has disappeared from the Lagrangian, while $\tilde{t}$ plays the role of $r$ in equation (\ref{eq:timelike L}). The only difference is an overall factor of $i$ in the non-diagonal terms, which does not affect the trace nor the determinant of the matrix $D$. Thus, the stability of perturbations that only depend on $z$ (and time) is determined by the same conditions we derived in the previous section, with mass parameters given by equations (\ref{eq:mu tilde}) and (\ref{eq:mt tilde}). It follows for instance  that for generic values of $\beta, \beta_1$, the theory has a high-momentum ghost. 

The presence of a ghost at high momenta is not exclusive of $z$ dependent perturbations, and also appears for perturbations that only depend on $x$ and $y$. Indeed,  setting $p_{\parallel}=0$ into the action (\ref{eq:spacelike L}) gives
\begin{eqnarray}
	\mathcal{S}=-\int \frac{d^4p}{(2\pi)^{4}} &\Big\{&\left[\beta \, \omega^2-\beta_1 \vec{p}_{\perp}\!^2 +\tilde{m}_u^2 \right]\tilde{u}^*\tilde{u}
	+\left[-\beta_1 \omega^2 \vec{p}_{\perp}\!^2 +
	\beta\vec{p}_{\perp}\!^4 +\vec{p}_{\perp}\!^2 m_v^2\right] \tilde{r}^*\tilde{r}
	 - \nonumber \\
	{}&&-2\beta H \vec{p}_{\perp}\!^2\left[\tilde{u}^*\tilde{r}+\tilde{r}^*\tilde{u}\right]
	-i\omega (\beta-\beta_1) \vec{p}_{\perp}\!^2(\tilde{r}^* \tilde{u}-\tilde{u}^* \tilde{r})+\nonumber \\ 
	&&+\left[-\beta_1 \omega^2+\beta_1 \vec{p}_{\perp}\!^2+ \tilde{m}_t^2\right]\tilde{t}^*\tilde{t}\Big\}.
\end{eqnarray}
The field $\tilde{t}$ has decoupled from the other two scalars,  whose action has again the form of (\ref{eq:timelike L}) (after canonically normalizing $\tilde{r}$.) Thus, as before, the theory has ghosts at sufficiently high values of  $\vec{p}\,^2_{\perp}$ for generic values of $\beta, \beta_1$.

For $\beta=0$ or $\beta_1=0$ these instabilities disappear if appropriate conditions are satisfied. These are listed in Tables \ref{table:b0 spacelike stability} and \ref{table:b1 spacelike stability}. As an application,  let us consider for instance Lagrangians with $\beta=0$, as in the ``new inflation" vector models of  \cite{Ford:1989me}, or as in the dark energy model of \cite{ArmendarizPicon:2004pm}. Using equations (\ref{eq:mv}), (\ref{eq:mu tilde}) and (\ref{eq:mt tilde}), and neglecting curvature terms  we find
\begin{eqnarray}
	\tilde{m}_t^2-\tilde{m}_u^2&\approx& 2V'+2V'' \bar{A}^2\\
	m_v^2-\tilde{m}_u^2&\approx&2V' \\
	m_v^2&\approx&V'.
\end{eqnarray}
It follows automatically from our analysis that all models with ``run-away" potentials with $V'<0$, in which the derivatives of the potential dominate over curvature terms,  contain ghosts in the scalar sector at high momenta, and are thus quantum-mechanically unstable \cite{Dubovsky}.
\begin{table}
\begin{tabular}{|c|c|c|}\hline
 {\bf Scalar} ($\beta=0$) & Low $p$ & High $p$ \\ \hline 
 Classical & $ \beta_1 \tilde{m}_t^2>0,\, \beta_1 m_v^2>0$ &  $\tilde{m}_t^2\tilde{m}_u^2<0,\, m_v^2\tilde{m}_u^2<0$ \\ \hline
Quantum & $\beta_1>0$ & $\beta_1>0,\, \tilde{m}_t^2-\tilde{m}_u^2>0,\, m_v^2-\tilde{m}_u^2>0$  \\ \hline
\end{tabular}
\caption{Stability conditions in the scalar sector for spacelike background fields and $\beta=0$.}\label{table:b0 spacelike stability}
\end{table}

\begin{table}
\begin{tabular}{|c|c|c|}\hline
 {\bf Scalar} ($\beta_1=0$) & Low $p$ & High $p$ \\ \hline 
 Classical & $\beta  \tilde{m}_u^2<0$ & $ \tilde{m}_r^2(\tilde{m}_u^2-4\beta H^2)<0,\, m_v^2(\tilde{m}_u^2-4\beta H^2)<0$ \\ \hline
Quantum & $\beta<0$ & $\tilde{m}_u^2-\tilde{m}_t^2-4\beta H^2>0,\, \tilde{m}_u^2-m_v^2-4\beta H^2>00$  \\ \hline
\end{tabular}
\caption{Stability conditions in the scalar sector for spacelike background fields and $\beta_1=0$.}\label{table:b1 spacelike stability}
\end{table}

Let us conclude by emphasizing that the  stability conditions we have derived in this Appendix are necessary but  not sufficient. Because we have not analyzed the stability of all Fourier modes, a violation of our conditions points to the instability of certain field modes, but their satisfaction does not imply that all the modes are free of instabilities.  
\end{appendix}


\begin{thebibliography}{99}
  
\bibitem{Will:1972zz}
  C.~M.~Will and K.~J.~Nordtvedt,
  ``Conservation Laws and Preferred Frames in Relativistic Gravity. I.
  Preferred-Frame Theories and an Extended PPN Formalism,''
  Astrophys.\ J.\  {\bf 177}, 757 (1972).
  
\bibitem{Hellings:1973zz}
  R.~W.~Hellings and K.~Nordtvedt,
  ``Vector-Metric Theory of Gravity,''
  Phys.\ Rev.\  D {\bf 7}, 3593 (1973).
  
  \bibitem{Ford:1989me}
  L.~H.~Ford,
  ``Inflation Driven by a Vector Field,''
  Phys.\ Rev.\  D {\bf 40}, 967 (1989).

  \bibitem{ArmendarizPicon:2004pm}
  C.~Armendariz-Picon,
  ``Could dark energy be vector-like?,''
  JCAP {\bf 0407}, 007 (2004)
  [arXiv:astro-ph/0405267].
  
   \bibitem {Wei:2006tn}
  H.~Wei and R.~G.~Cai,
  ``Interacting vector-like dark energy, the first and second cosmological
  coincidence problems,''
  Phys.\ Rev.\  D {\bf 73}, 083002 (2006)
  [arXiv:astro-ph/0603052].
  
  \bibitem{Kanno:2006ty}
  S.~Kanno and J.~Soda,
  ``Lorentz violating inflation,''
  Phys.\ Rev.\  D {\bf 74}, 063505 (2006)
  [arXiv:hep-th/0604192].
  
  \bibitem{Novello:2006ng}
  M.~Novello, E.~Goulart, J.~M.~Salim and S.~E.~Perez Bergliaffa,
  ``Cosmological effects of nonlinear electrodynamics,''
  Class.\ Quant.\ Grav.\  {\bf 24}, 3021 (2007)
  [arXiv:gr-qc/0610043].
  
\bibitem{Boehmer:2007qa}
  C.~G.~Boehmer and T.~Harko,
  ``Dark energy as a massive vector field,''
  Eur.\ Phys.\ J.\  C {\bf 50}, 423 (2007)
  [arXiv:gr-qc/0701029].
  
  \bibitem{Koivisto:2007bp}
  T.~Koivisto and D.~F.~Mota,
  ``Accelerating Cosmologies with an Anisotropic Equation of State,''
  Astrophys.\ J.\  {\bf 679}, 1 (2008)
  [arXiv:0707.0279 [astro-ph]].

\bibitem{Jimenez:2008au}
  J.~B.~Jimenez and A.~L.~Maroto,
  ``A cosmic vector for dark energy,''
  Phys.\ Rev.\  D {\bf 78}, 063005 (2008)
  [arXiv:0801.1486 [astro-ph]].
  
  \bibitem{Koivisto:2008ig}
  T.~Koivisto and D.~F.~Mota,
  ``Anisotropic Dark Energy: Dynamics of Background and Perturbations,''
  JCAP {\bf 0806}, 018 (2008)
  [arXiv:0801.3676 [astro-ph]].
  
  \bibitem{Golovnev:2008cf}
  A.~Golovnev, V.~Mukhanov and V.~Vanchurin,
  ``Vector Inflation,''
  JCAP {\bf 0806}, 009 (2008)
  [arXiv:0802.2068 [astro-ph]].
  
\bibitem{Koivisto:2008xf}
  T.~S.~Koivisto and D.~F.~Mota,
  ``Vector Field Models of Inflation and Dark Energy,''
  JCAP {\bf 0808}, 021 (2008)
  [arXiv:0805.4229 [astro-ph]].
  
  \bibitem{Yokoyama:2008xw}
  S.~Yokoyama and J.~Soda,
  ``Primordial statistical anisotropy generated at the end of inflation,''
  JCAP {\bf 0808}, 005 (2008)
  [arXiv:0805.4265 [astro-ph]].
  
  \bibitem{Kanno:2008gn}
  S.~Kanno, M.~Kimura, J.~Soda and S.~Yokoyama,
  ``Anisotropic Inflation from Vector Impurity,''
  JCAP {\bf 0808}, 034 (2008)
  [arXiv:0806.2422 [hep-ph]].
  
  \bibitem{Watanabe:2009ct}
  M.~a.~Watanabe, S.~Kanno and J.~Soda,
  ``Hairy Inflation,''
  arXiv:0902.2833 [hep-th].
  
  \bibitem{Koh:2009ne}
  S.~Koh,
  ``Vector Field and Inflation,''
  arXiv:0902.3904 [hep-th].
 
 \bibitem {Jacobson:2000xp}
  T.~Jacobson and D.~Mattingly,
  ``Gravity with a dynamical preferred frame,''
  Phys.\ Rev.\  D {\bf 64}, 024028 (2001)
  [arXiv:gr-qc/0007031].
   
  
 \bibitem{Gasperini:1987nq}
  M.~Gasperini,
  ``Singularity Prevention and Broken Lorentz
   Symmetry,"
  Class.\ Quant.\ Grav.\  {\bf 4}, 485 (1987).
 
 \bibitem{Jacobson:2008aj}
  T.~Jacobson,
  ``Einstein-aether gravity: a status report,''
  PoS {\bf QG-PH}, 020 (2007)
  [arXiv:0801.1547 [gr-qc]].

\bibitem{Kostelecky:1988zi}
  V.~A.~Kostelecky and S.~Samuel,
  ``Spontaneous Breaking of Lorentz Symmetry in String Theory,''
  Phys.\ Rev.\  D {\bf 39}, 683 (1989).
  
 \bibitem{Horava:2009uw}
  P.~Horava,
  ``Quantum Gravity at a Lifshitz Point,''
  Phys.\ Rev.\  D {\bf 79}, 084008 (2009)
  [arXiv:0901.3775 [hep-th]].


\bibitem{Collins:2004bp}
  J.~Collins, A.~Perez, D.~Sudarsky, L.~Urrutia and H.~Vucetich,
  ``Lorentz invariance and quantum gravity: an additional fine-tuning problem?,''
  Phys.\ Rev.\ Lett.\  {\bf 93}, 191301 (2004)
  [arXiv:gr-qc/0403053]. 
  
 \bibitem{Charmousis:2009tc}
  C.~Charmousis, G.~Niz, A.~Padilla and P.~M.~Saffin,
  ``Strong coupling in Horava gravity,''
  JHEP {\bf 0908}, 070 (2009)
  [arXiv:0905.2579 [hep-th]]. 
 
  \bibitem{Blas:2009yd}
  D.~Blas, O.~Pujolas and S.~Sibiryakov,
  ``On the Extra Mode and Inconsistency of Horava Gravity,''
  JHEP {\bf 0910}, 029 (2009)
  [arXiv:0906.3046 [hep-th]].
  
  \bibitem{ArkaniHamed:2003uy}
  N.~Arkani-Hamed, H.~C.~Cheng, M.~A.~Luty and S.~Mukohyama,
  ``Ghost condensation and a consistent infrared modification of gravity,''
  JHEP {\bf 0405}, 074 (2004)
  [arXiv:hep-th/0312099].
  
  \bibitem {Bluhm:2004ep}
  R.~Bluhm and V.A.~Kostelecky,
  ``Spontaneous Lorentz violation, Nambu-Goldstone modes, and gravity,''
  Phys.\ Rev.\  D {\bf 71}, 065008 (2005)
  [arXiv:hep-th/0412320]. 
  
  \bibitem {Bailey:2006fd}
  Q.G.~Bailey and V.A.~Kostelecky,
  ``Signals for Lorentz violation in post-Newtonian gravity,''
  Phys.\ Rev.\  D {\bf 74}, 045001 (2006)
  [arXiv:gr-qc/0603030].
  
  \bibitem {Bluhm:2007bd}
  R.~Bluhm, S-H.~Fung and V.A.~Kostelecky,
  ``Spontaneous Lorentz and Diffeomorphism Violation, Massive Modes, and Gravity,''
  Phys.\ Rev.\  D {\bf 77}, 065020 (2008)
  [arXiv:0712.4119 [hep-th]].
  
  \bibitem {Kostelecky:2009zr}
  V.A.~Kostelecky and R.~Potting,
  ``Gravity from spontaneous Lorentz violation,''
  Phys.\ Rev.\  D {\bf 79}, 065018 (2009)
  [arXiv:0901.0662 [gr-qc]].
  
  
    
\bibitem {Weinberg:gauge}
  S.~Weinberg,
  ``The Quantum theory of fields. Vol. 1: Foundations,'' Section 5.9, 
{\it  Cambridge, UK: Univ. Pr. (1995) 609 p}.

\bibitem{Colladay:1998fq}
  D.~Colladay and V.~A.~Kostelecky,
  ``Lorentz-violating extension of the standard model,''
  Phys.\ Rev.\  D {\bf 58}, 116002 (1998)
  [arXiv:hep-ph/9809521].

\bibitem{Kostelecky:2008ts}
  V.~A.~Kostelecky and N.~Russell,
  ``Data Tables for Lorentz and CPT Violation,''
  arXiv:0801.0287 [hep-ph].
  
  \bibitem{Mattingly:2005re}
  D.~Mattingly,
  ``Modern tests of Lorentz invariance,''
  Living Rev.\ Rel.\  {\bf 8}, 5 (2005)
  [arXiv:gr-qc/0502097].

\bibitem{Burgess:2007pt}
  C.~P.~Burgess,
  ``Introduction to effective field theory,''
  Ann.\ Rev.\ Nucl.\ Part.\ Sci.\  {\bf 57}, 329 (2007)
  [arXiv:hep-th/0701053].

 \bibitem{Will:1981cz}
  C.~M.~Will,
  ``Theory And Experiment In Gravitational Physics,''
{\it  Cambridge, Uk: Univ. Pr. ( 1981) 342p}.

\bibitem{Carroll:2004ai}
  S.~M.~Carroll and E.~A.~Lim,
  ``Lorentz-violating vector fields slow the universe down,''
  Phys.\ Rev.\  D {\bf 70}, 123525 (2004)
  [arXiv:hep-th/0407149].

\bibitem{Bjorken:1963vg}
  J.~D.~Bjorken,
  Annals Phys.\  {\bf 24}, 174 (1963).

 \bibitem{Kraus:2002sa}
  P.~Kraus and E.~T.~Tomboulis,
  ``Photons and gravitons as Goldstone bosons, and the cosmological
  constant,''
  Phys.\ Rev.\  D {\bf 66}, 045015 (2002)
  [arXiv:hep-th/0203221].
      
\bibitem{Eling:2003rd}
  C.~Eling and T.~Jacobson,
  ``Static post-Newtonian equivalence of GR and gravity with a dynamical preferred frame,''
  Phys.\ Rev.\  D {\bf 69}, 064005 (2004)
  [arXiv:gr-qc/0310044].
  
 \bibitem{Foster:2005dk}
  B.~Z.~Foster and T.~Jacobson,
  ``Post-Newtonian parameters and constraints on Einstein-aether theory,''
  Phys.\ Rev.\  D {\bf 73}, 064015 (2006)
  [arXiv:gr-qc/0509083]. 
  
\bibitem{Bertotti:2003rm}
  B.~Bertotti, L.~Iess and P.~Tortora,
  ``A test of general relativity using radio links with the Cassini
  spacecraft,''
  Nature {\bf 425}, 374 (2003).  
  
\bibitem{Zlosnik:2006zu}
  T.~G.~Zlosnik, P.~G.~Ferreira and G.~D.~Starkman,
  ``Modifying gravity with the Aether: an alternative to Dark Matter,''
  Phys.\ Rev.\  D {\bf 75}, 044017 (2007)
  [arXiv:astro-ph/0607411].
  
\bibitem{Milgrom:1983ca}
 M.~Milgrom, ``A Modification Of The Newtonian Dynamics As A Possible Alternative To The Hidden Mass Hypothesis,''
  Astrophys.\ J.\  {\bf 270}, 365 (1983).

\bibitem{Bekenstein:2004ne}
  J.~D.~Bekenstein,
  ``Relativistic gravitation theory for the MOND paradigm,''
  Phys.\ Rev.\  D {\bf 70}, 083509 (2004)
  [Erratum-ibid.\  D {\bf 71}, 069901 (2005)]
  [arXiv:astro-ph/0403694].  
  
 \bibitem{Dvali:2007kt}
  G.~Dvali, S.~Hofmann and J.~Khoury,
  ``Degravitation of the cosmological constant and graviton width,''
  Phys.\ Rev.\  D {\bf 76}, 084006 (2007)
  [arXiv:hep-th/0703027]. 
    
\bibitem{Lim:2004js}
  E.~A.~Lim,
  ``Can we see Lorentz-violating vector fields in the CMB?,''
  Phys.\ Rev.\  D {\bf 71}, 063504 (2005)
  [arXiv:astro-ph/0407437].
  
 \bibitem{Carroll:2008em}
  S.~M.~Carroll, T.~R.~Dulaney, M.~I.~Gresham and H.~Tam,
  ``Instabilities in the Aether,''
  arXiv:0812.1049 [hep-th].
  
\bibitem{Jimenez:2008sq}
  J.~B.~Jimenez and A.~L.~Maroto,
  ``Viability of vector-tensor theories of gravity,''
  arXiv:0811.0784 [astro-ph].
  
  \bibitem{Himmetoglu:2008zp}
  B.~Himmetoglu, C.~R.~Contaldi and M.~Peloso,
  ``Instability of anisotropic cosmological solutions supported by vector fields,''
  arXiv:0809.2779 [astro-ph].
  
  \bibitem{Himmetoglu:2008hx}
  B.~Himmetoglu, C.~R.~Contaldi and M.~Peloso,
  ``Instability of the ACW model, and problems with massive vectors during
  inflation,''
  arXiv:0812.1231 [astro-ph].
  
  \bibitem{Himmetoglu:2009qi}
  B.~Himmetoglu, C.~R.~Contaldi and M.~Peloso,
  ``Ghost instabilities of cosmological models with vector fields nonminimally
  coupled to the curvature,''
  arXiv:0909.3524 [astro-ph.CO].
  
  \bibitem{Dubovsky:2004sg}
  S.~L.~Dubovsky,
  ``Phases of massive gravity,''
  JHEP {\bf 0410}, 076 (2004)
  [arXiv:hep-th/0409124].
  
 \bibitem{Henneaux:1992ig}
  M.~Henneaux and C.~Teitelboim,
  ``Quantization of gauge systems,''
{\it  Princeton, USA: Univ. Pr. (1992) 520 p}

 
\bibitem{Mukhanov:2007zz}
  V.~Mukhanov and S.~Winitzki,
  ``Introduction to quantum effects in gravity,''
{\it  Cambridge, UK: Cambridge Univ. Pr. (2007) 273 p}

  \bibitem{Carroll:2003st}
  S.~M.~Carroll, M.~Hoffman and M.~Trodden,
  ``Can the dark energy equation-of-state parameter w be less than -1?,''
  Phys.\ Rev.\  D {\bf 68}, 023509 (2003)
  [arXiv:astro-ph/0301273].
  
 \bibitem {Cline:2003gs}
  J.~M.~Cline, S.~Jeon and G.~D.~Moore,
  ``The phantom menaced: Constraints on low-energy effective ghosts,''
  Phys.\ Rev.\  D {\bf 70}, 043543 (2004)
  [arXiv:hep-ph/0311312].

\bibitem{ArmendarizPicon:2008yv}
  C.~Armendariz-Picon, M.~Fontanini, R.~Penco and M.~Trodden,
  ``Where does Cosmological Perturbation Theory Break Down?,''
  arXiv:0805.0114 [hep-th].

\bibitem{Woodard:2006nt}
  R.~P.~Woodard,
  ``Avoiding dark energy with 1/R modifications of gravity,''
  Lect.\ Notes Phys.\  {\bf 720}, 403 (2007)
  [arXiv:astro-ph/0601672].

  
\bibitem{Weinberg:Rxi}
  S.~Weinberg,
  ``The quantum theory of fields. Vol. 2: Modern applications,'' Section 15.2, 
{\it  Cambridge, UK: Univ. Pr. (1996) 489 p}

\bibitem{Dvali:2005nt}
  G.~Dvali, M.~Papucci and M.~D.~Schwartz,
  ``Infrared Lorentz violation and slowly instantaneous electricity,''
  Phys.\ Rev.\ Lett.\  {\bf 94}, 191602 (2005)
  [arXiv:hep-th/0501157].
  
 \bibitem{Gabadadze:2004iv}
  G.~Gabadadze and L.~Grisa,
  ``Lorentz-violating massive gauge and gravitational fields,''
  Phys.\ Lett.\  B {\bf 617}, 124 (2005)
  [arXiv:hep-th/0412332].
  
  \bibitem{Dubovsky:2005xd}
  S.~Dubovsky, T.~Gregoire, A.~Nicolis and R.~Rattazzi,
  ``Null energy condition and superluminal propagation,''
  JHEP {\bf 0603}, 025 (2006)
  [arXiv:hep-th/0512260].

   
  \bibitem{ArmendarizPicon:2005nz}
  C.~Armendariz-Picon and E.~A.~Lim,
  ``Haloes of k-essence,''
  JCAP {\bf 0508}, 007 (2005)
  [arXiv:astro-ph/0505207].
  
  \bibitem{Adams:2006sv}
  A.~Adams, N.~Arkani-Hamed, S.~Dubovsky, A.~Nicolis and R.~Rattazzi,
  ``Causality, analyticity and an IR obstruction to UV completion,''
  JHEP {\bf 0610}, 014 (2006)
  [arXiv:hep-th/0602178].
  
 \bibitem {Bruneton:2006gf}
  J.~P.~Bruneton,
  ``On causality and superluminal behavior in classical field theories.
  Applications to k-essence theories and MOND-like theories of gravity,''
  Phys.\ Rev.\  D {\bf 75}, 085013 (2007)
  [arXiv:gr-qc/0607055].

\bibitem{Babichev:2007dw}
  E.~Babichev, V.~Mukhanov and A.~Vikman,
  ``k-Essence, superluminal propagation, causality and emergent geometry,''
  JHEP {\bf 0802}, 101 (2008)
  [arXiv:0708.0561 [hep-th]].
  
  \bibitem{Dubovsky:2008bd}
  S.~Dubovsky and S.~Sibiryakov,
  ``Superluminal Travel Made Possible (in two dimensions),''
  JHEP {\bf 0812}, 092 (2008)
  [arXiv:0806.1534 [hep-th]].
  
 \bibitem{Dubovsky}
The existence of instabilities in vector field theories with run-away potentials was suggested long ago to one of us by S. Dubovsky in a private communication. 

\end{thebibliography}
\end{document}